\documentclass[twocolumn,twocolappendix]{aastex701}

\usepackage{amsmath}
\usepackage{gensymb}

\begin{document}

\title{Updated in-flight calibration of the Hayabusa2/NIRS3 spectrometer: new global near-infrared photometric properties of asteroid (162173) Ryugu}

\author[orcid=0000-0003-4335-029X]{Antonin Wargnier}
\affiliation{Institute of Space and Astronautical Science (ISAS), Japan Aerospace Exploration Agency (JAXA), Sagamihara, Kanagawa 2525210, Japan}
\email[show]{wargnier.antonin@jaxa.jp}  

\author[orcid=0000-0002-6142-9842]{Eri Tatsumi} 
\affiliation{Instituto de Astrofísica de Canarias (IAC), La Laguna, Tenerife, Spain}
\affiliation{Department of Astrophysics, University of La Laguna, La Laguna, Tenerife, Spain}
\affiliation{Department of Earth and Planetary Science, The University of Tokyo, Bunkyo, Tokyo, Japan}
\email{eri.tatsumi@iac.es}

\author[orcid=0000-0001-6160-9360]{Koki Yumoto}
\affiliation{Institute of Space and Astronautical Science (ISAS), Japan Aerospace Exploration Agency (JAXA), Sagamihara, Kanagawa 2525210, Japan}
\affiliation{LIRA, Observatoire de Paris, Université PSL, Sorbonne Université, Université Paris Cité, CNRS, CY Cergy Paris Université, 5 place Jules Janssen, Meudon, 92195, France}
\email{yumoto.koki@jaxa.jp}

\author[]{Mayumi Ichikawa}
\affiliation{Institute of Space and Astronautical Science (ISAS), Japan Aerospace Exploration Agency (JAXA), Sagamihara, Kanagawa 2525210, Japan}
\email{ichikawa.mayumi@jaxa.jp}

\author[orcid=0000-0002-7137-4849]{Shin-ya Murakami}
\affiliation{Institute of Space and Astronautical Science (ISAS), Japan Aerospace Exploration Agency (JAXA), Sagamihara, Kanagawa 2525210, Japan}
\email{murakami.shinya@jaxa.jp}

\author[]{Yuichiro Nagai}
\affiliation{Institute of Space and Astronautical Science (ISAS), Japan Aerospace Exploration Agency (JAXA), Sagamihara, Kanagawa 2525210, Japan}
\email{nagai.yuuichiroh@jaxa.jp}

\author[]{Kazuhiro Honda}
\affiliation{Institute of Space and Astronautical Science (ISAS), Japan Aerospace Exploration Agency (JAXA), Sagamihara, Kanagawa 2525210, Japan}
\email{honda.kazuhiro@jaxa.jp}

\author[]{Yasuhiro Yokota}
\affiliation{Institute of Science Tokyo, Ookayama, Tokyo, Japan}
\email{yokota@planeta.sci.isas.jaxa.jp}

\author[orcid=0000-0002-1060-3986]{Toru Kouyama}
\affiliation{Artificial Intelligence Research Center, National Institute of Advanced Industrial Science and Technology, 2-4-7 Aomi, Koto-ku, Tokyo 135-0064, Japan}
\email{t.kouyama@aist.go.jp}

\author[orcid=0000-0002-4613-7956]{Takahiro Iwata}
\affiliation{Institute of Space and Astronautical Science (ISAS), Japan Aerospace Exploration Agency (JAXA), Sagamihara, Kanagawa 2525210, Japan}
\affiliation{The Graduate University for Advanced Studies, SOKENDAI, Sagamihara, Kanagawa, 252-5210, Japan}
\email{iwata.takahiro@jaxa.jp}

\author[orcid=0000-0003-1091-3041]{Moe Matsuoka}
\affiliation{Geological Survey of Japan, National Institute of Advanced Industrial Science and Technology, Tsukuba, Japan}
\email{moe.matsuoka@aist.go.jp}

\author[orcid=0000-0002-4874-0417]{Satoshi Tanaka}
\affiliation{Institute of Space and Astronautical Science (ISAS), Japan Aerospace Exploration Agency (JAXA), Sagamihara, Kanagawa 2525210, Japan}
\affiliation{The Graduate University for Advanced Studies, SOKENDAI, Sagamihara, Kanagawa, 252-5210, Japan}
\email{tanaka@planeta.sci.isas.jaxa.jp}

\author[orcid=0000-0002-4809-7492]{Kohei Kitazato}
\affiliation{Aizu Research Cluster for Space Science, University of Aizu, Aizu-Wakamatsu 965-8580, Japan}
\email{kitazato@u-aizu.ac.jp}

\begin{abstract}
   The Near InfraRed Spectrometer (NIRS3) onboard Hayabusa2 observed asteroid (162173) Ryugu, searching especially for signatures of water-bearing minerals. However, during the mission, absolute reflectance measured by the Telescopic Optical Navigation Camera (ONC-T) and NIRS3 showed a systematic offset. Additionally, touchdown operations have modified the instrumental response, which has made post-second-touchdown data less reliable. Correction of these issues is crucial to interpret the subtle spectroscopic differences across the surface, better understand the origin and evolution of carbonaceous asteroids, and support the Hayabusa2$\sharp$ mission's future observations of the asteroids (98943) Torifune and 1998 KY26.
   We used Ryugu observations at different mission phases for relative calibrations and lunar observations for the absolute calibration. Moon spectra acquired during the first Earth swing-by were compared to simulated ones from photometric models based on lunar orbiter and ground-based data. We then applied the updated calibration to the full Hayabusa2 proximity phase dataset, enabling disk-resolved global and regional photometric analyses.
   We showed that the data obtained following the first and second touchdowns required downscaled corrections of $5.4 \pm 1.8 $\% and $10.3 \pm 0.4 $\% at 2 $\mu m$, respectively, while the absolute calibration revealed that NIRS3 reflectance was underestimated by about 8\% compared with the original calibration. Our new photometric parameters are consistent with previous studies and indicate a weak east-west dichotomy in phase ratio and phase reddening, pointing to higher surface roughness and higher abundance of fine-grained regolith in Ryugu eastern hemisphere. 
\end{abstract}



\section{Introduction}
The JAXA Hayabusa2 mission, devoted to the study of the carbonaceous near-Earth asteroid (162173) Ryugu, characterized for the first time the geology, the composition, and the physical properties of this type of asteroid. Among the suite of instruments, the Near-InfraRed Spectrometer (NIRS3, \citealt{Iwata_2017}) provides measurements of the composition and constraints on the water content of Ryugu \citep{Praet_2021}. NIRS3 revealed that the surface of Ryugu is particularly dark in the near-infrared, red-sloped, and present a weak and narrow 2.72 $\mu m$ OH band -- linked to the presence of phyllosilicates -- across the entire surface \citep{Kitazato_2019}. \par
After successfully bringing back sample from Ryugu in 2020, the Hayabusa2 spacecraft is now in its extended mission phase as Hayabusa2$\sharp$ \citep{Tsuda_2025}, and on its way to two new asteroids, with a flyby on 5 July 2026 of the S-type (98943) Torifune \citep{Geem_2023, Popescu_2025}, and a rendezvous with the small fast-rotator 1998 KY26 in 2031 \citep{Bolin_2025, Beniyama_2025, Santana-Ros_2025}. \par
After the two touchdown (TD) operations of the Hayabusa2 spacecraft on 21 February and 11 July 2019, the instrumental response of the NIRS3 instrument was altered likely due to contamination by dust particles, and the radiometric calibration was modified accordingly, thanks to the measurements from the radiometric (RAD) calibration lamp \citep{Kitazato_2021}. However, because the optical path of the RAD calibration lamp is not identical to the optical path of the science beam, taking into account the signal from the RAD calibration lamp only, may have lead to an over or undercorrection of the radiometric calibration coefficient (RCC). This is further supported from ONC-T observations, which showed that the degrees of dust contamination inferred from calibration lamps and star observations were not consistent \citep{Kouyama_2021}. In fact, thorough analysis of all the Ryugu data obtained during the proximity phase showed systematic residual variations between data obtained during mission phases, indicating the need for further refinement of the radiometric calibration (Fig. \ref{fig:phase_curved_by_phases}). \par
In the context of the data archiving of the Ryugu data and of the extended mission phase, the need for a precise calibration is important because it ensures the accuracy of measurements and the reliability of the results.  In addition, this new calibration enables the definition of a new global photometric properties for the Ryugu spectra acquired by the NIRS3 instrument using the full dataset acquired during the proximity phase. Previous studies have derived photometric properties from partial dataset, systematically excluding data obtained after TD2 \citep{Pilorget_2021, Domingue_2021}. However, data acquired after TD2 are particularly valuable for photometric studies because they cover the opposition effect with very low phase angle observations, but also the larger phase angle from 35$\degree$ to almost 50$\degree$.\par
In this work, we first provide an updated calibration of NIRS3 based on the observations of the Moon (Sect. \ref{sect:moon_absolute_calib}) and Ryugu (Sect. \ref{sect:relative_calib}). This update calibration is important not only
for the nominal mission data analyses but also for the future
extended mission data. Then, we determine new photometric properties from this newly calibrated data and discuss the results (Sect. \ref{sec:disk_resolved}).

\section{The NIRS3 instrument and dataset}
NIRS3 is a near-infrared spectrometer with an effective wavelength range from 1.8 $\mu m$ to 3.2 $\mu m$ \citep{Iwata_2017, Kitazato_2021}. The NIRS3 detector consists in 128 InAs photodiodes, which give an average spectral sampling of 18 nm per channel. The raw science data -- corresponding to processing level "L1A" -- were converted to calibrated science "L2C" data by considering the calibration data, such as wavelength-channels conversion, electronic offset or the radiometric calibration coefficient (RCC). These calibrated science data were converted into radiance factor (I/F) unit for each wavelength value $\lambda$ corresponding to the center wavelength of each NIRS3 channel by considering the following equation:
\begin{equation}
    \left.\frac{I}{F}\right|_{\lambda} =  \frac{\pi d^{2}}{F_\lambda} \cdot \left(DN_{\text{mean},\lambda} - DN_{\text{offset},\lambda}\right) \cdot RCC_{\lambda}
\end{equation}
where $d$ is the Sun-target distance in au, $F_{\lambda}$ is the solar spectral irradiance at 1 au [W\,m$^{-2}$\,nm$^{-1}$], $DN_{\text{mean}, \lambda}$ and $DN_{\text{offset}, \lambda}$ are the raw science signals in digital counts per second [DN\,s$^{-1}$] and the electronic offsets in DN\,s$^{-1}$, respectively. $RCC_\lambda$ corresponds to the radiometric calibration coefficient for each wavelength channel $\lambda$, allowing to convert digital counts to radiance unit, given in (W\,m$^{-2}$\,nm$^{-1}$\,sr$^{-1}$)/(DN\,s$^{-1}$). The $RCC_{\lambda}$ coefficients were obtained through the RAD internal calibration lamp during cruise phase and updated afterward during other mission phases \citep{Kitazato_2021}. \newline
After obtaining radiance factor spectra, the thermal emission from the target was removed, leading to the creation of "L2D" data products \citep{Ichikawa_2026a, Ichikawa_2026b}. The thermal radiance was estimated assuming a single-temperature emission described by Planck's law modulated by a wavelength-dependent emissivity $\varepsilon(\lambda)$ constrained through Kirchoff's law of thermal radiation. The observation geometry was calculated based on the SPICE kernels and the latest version of the highest resolution SPC shape model with 1 meter-size facets (ryugu\_shape\_spc\_3m\_v20200323.bds, \citealt{Murakami_2025}). \par
In this study, we utilized L2D NIRS3 data and corresponding geometry obtained during the proximity phase from June 2018 to November 2019 which are available through NASA’s Planetary Data System (PDS) and JAXA’s Data ARchive and Transmission System (DARTS).

\section{Updated in-flight calibration of NIRS3} 
\subsection{NIRS3 absolute calibration} \label{sect:moon_absolute_calib}
\begin{figure}
\centering
  \resizebox{\hsize}{!}{\includegraphics{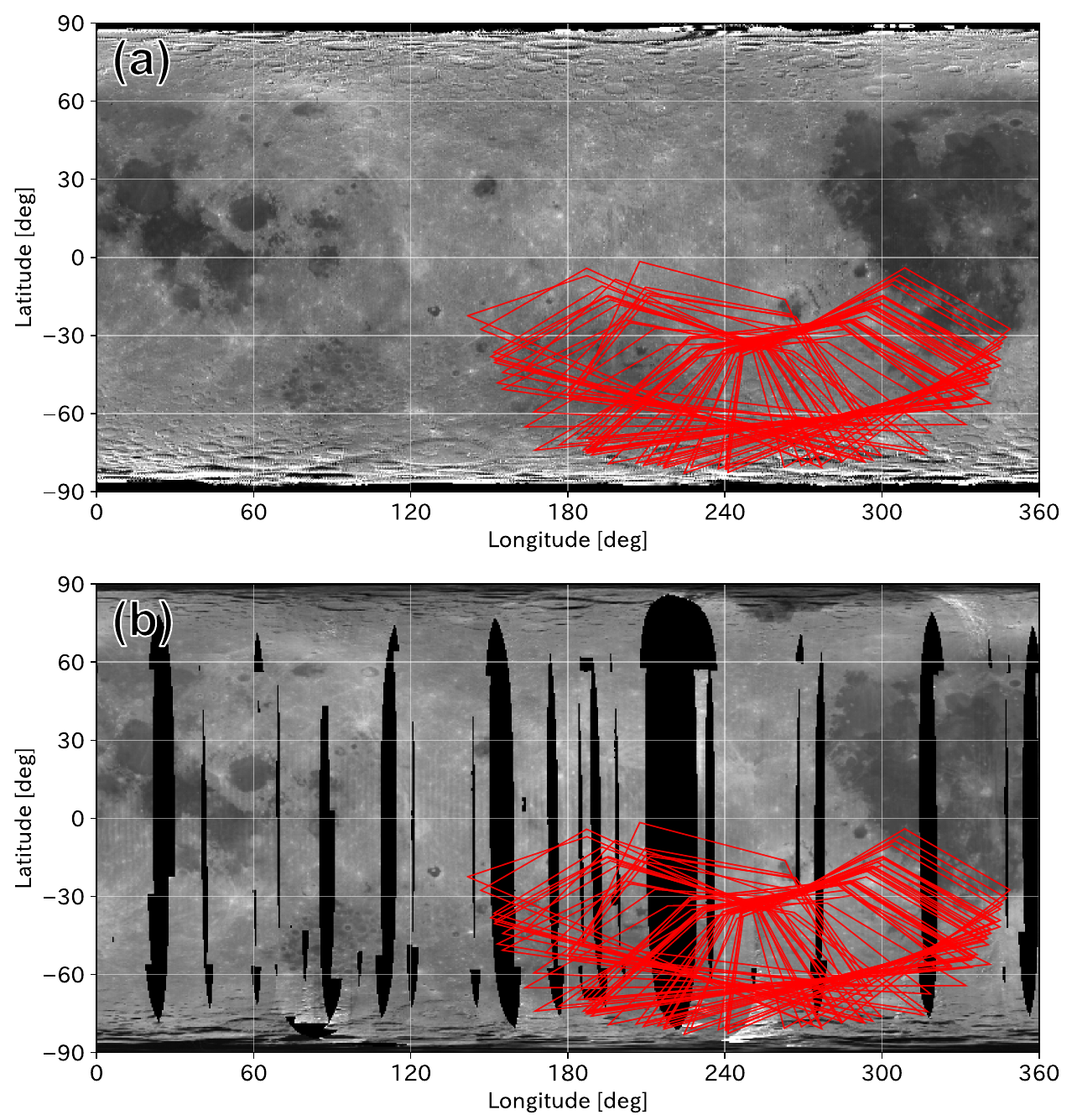}}
  \caption{Map of the Moon from (a) SP data \citep{Yokota_2011, Yokota_2012} (b) M$^3$ data, with projected footprints of NIRS3 (in red) during Hayabusa2 Earth swing-by of December 5, 2015.}
  \label{fig:moon_nirs3_footprint}
\end{figure}

The Moon is a particularly suitable target for in-flight calibration because it is an extended object, with a well-characterized surface, and stable over time \citep{Kieffer_1997}. We used Moon observations acquired by the NIRS3 spectrometer on December 5, 2015 during the Earth swing-by (Fig. \ref{fig:moon_nirs3_footprint} and Table \ref{tab:NIRS3_lunar_observations_properties}). The calibration on the Moon is performed with Kaguya/Spectral Profiler (SP) data following the method developed in \cite{Kouyama_2016} and Chandrayaan-1/Moon Mineralogy Mapper (M$^3$) data. The method with the SP data was also applied for the Hayabusa2 multi-band camera ONC-T \citep{Yumoto_2024}. We briefly summarized the method in the following. \par

\begin{table}[]
    \centering
    \caption{Observation characteristics and geometries of the NIRS3 lunar spectra. 
    }
    \begin{tabular}{cc}
    \hline
    \hline
       & Lunar NIRS3 spectra characteristics\\
    \hline
       Obs. date & 5 December 2015\\
       r$_h$\tablenotemark{a} & $0.984$ \\
       $\delta$\tablenotemark{a} & $\sim769\,000$\\
       $\alpha_M$\tablenotemark{a} & 59.3$\degree$\\
       Sub-S/C lat., lon. & 56.4$\degree$S,263.5$\degree$E\\
       Sub-solar lat., lon. & 1.5$\degree$N, 247.7$\degree$E\\
       No. spectra & 296\\
       No. used spectra\tablenotemark{b} & 45\\
    \hline
    \end{tabular}
    \label{tab:NIRS3_lunar_observations_properties}
    \tablenotetext{a}{
    r$_h$ is the Moon-Sun distance [au], $\delta$ is the S/C-Moon distance [km], $\alpha_M$ is the solar phase angle at the Moon center [deg].}
    \tablenotetext{b}{
    Some observations do not cover the full lunar disk and were hence not considered in this work. Spectra acquired with $i > 70\degree$ or $e > 70\degree$ were also discarded.}
\end{table}

\subsubsection{The SP model}
We simulated the Moon spectra within the NIRS3 footprints over the wavelength range 512 -- 2053 nm from the photometrically-corrected at the standard geometry (incidence angle $i_0 = 30\degree$, emission angle $e_0 = 0\degree$, phase angle $\alpha_0 = 30\degree$) radiance factor map derived from SP data (SP model; \citealt{Yokota_2011, Yokota_2012}). The simulated reflectance at the geometry $(i,e,\alpha)$ is given by:
\begin{equation}
    r_{\text{sim}}(\lambda,i,e,\alpha) = r_{\text{norm}}(\lambda,i_0,e_0,\alpha_0) \cdot \frac{r_{\text{model}}(\lambda,i,e,\alpha)}{r_{\text{model}}(\lambda,i_0,e_0,\alpha_0)} 
\end{equation}
In this equation, $r_{norm}(\lambda,i_0,e_0,\alpha_0)$ corresponds to the SP radiance factor spectra at the standard geometry which is computed as the average radiance factor within the projected NIRS3 footprints (Fig. \ref{fig:moon_nirs3_footprint}). $r_{\text{model}}$ represents the modeled radiance factor from the SP model (Appendix \ref{appendix: models}) and its associated wavelength-dependent parameters prepared by \cite{Yokota_2011, Yokota_2012}. Three different SP model parameter sets were defined according to the reflectance value of the surface materials, which then allows a distinction to be made between mare and highlands. We computed $r_{\text{model}}$ for both standard geometry ($i_0 = 30\degree, e_0 = 0\degree, \alpha_0 = 30\degree$), and NIRS3 geometry ($i,e,\alpha$). We calculated the geometries from the SPICE kernels considering instrument alignment, and spacecraft position and attitude. For NIRS3, the incidence and emission angles were determined by dividing the field-of-view into $3\times3$ sub-regions. These angles were evaluated at the center of each sub-region, and the final value corresponds to the median of the nine measurements. As phase angle variations are expected to be minor, only the value at the boresight vector was calculated for this angle. \par
The SP data were absolutely calibrated with laboratory measurements of the Apollo samples (e.g., \citealt{Matsunaga_2008}). Hence, for our NIRS3 lunar calibration, we needed to consider an independent lunar radiance observation data to convert SP data to the actual lunar surface radiance. For this purpose, \cite{Kouyama_2016} proposed to use the Robotic Lunar Observatory (ROLO) data, with which we can derive the ROLO correction factor $p(\lambda)$ as follows:
\begin{equation}
    p(\lambda) = a_0 + a_1 \lambda + a_2 \lambda^2 + a_3 \lambda^3,
\end{equation}
where $a_0$, $a_1$, $a_2$, and $a_3$ are the fitting coefficients of ratios between the SP and ROLO models. While the ROLO correction factor fitting coefficients were determined by \cite{Kouyama_2016} and updated by \cite{Yumoto_2024}, we needed, for this work, to extend this ROLO correction factor to cover the wavelengths from 1620 to 2053 nm.

\subsubsection{ROLO correction}
The ROLO albedo in the band $k$, $A_{k}$, is given by the following equation:
\begin{align}
    \ln A_k &= \sum_{i=0}^3 z_{ik} \alpha^{i} + \sum_{j=1}^3 b_{jk} \Phi^{2j-1} + c_1 \theta + c_2 \varphi + c_3 \Phi \theta  + c_4 \Phi\varphi \nonumber \\ &+ d_{1k} e^{-\alpha/p_1} + d_{2k} e^{-\alpha/p_2} + d_{3k} \cos{\left[{\frac{\alpha-p_3}{p_4}}\right]},
\end{align}
where $z_{ik}$, $b_{jk}$, $c_{1}$, $c_{2}$, $c_{3}$, $c_{4}$, $d_{1k}$, $d_{2k}$, $d_{3k}$, $p_{1}$, $p_{2}$, $p_{3}$, and $p_{4}$ are coefficients given in \cite{Kieffer_2005}. $\alpha$ is the phase angle, $\Phi$ is the sub-solar longitude, $\varphi$ and $\theta$ are the longitude and latitude of the sub-observer point on the Moon, respectively. \newline
We then computed the Moon irradiance in the band $k$, $I_k$, using:
\begin{equation}
    I_{k} = \frac{A_{k} \Omega_{M} E_{k}}{\pi}
\end{equation}
where $\Omega_M$ is the solid angle of the Moon, which is equal to $6.4177 \times 10^{-5}$ sr. $E_k$ describes the solar irradiance integrated over the bandpass of the $k$ band. The lunar irradiance is finally converted to irradiance at standard distances: $I_{0,k} = I_{k} \cdot \left(\frac{D_{Sun-Moon}}{1 \text{ au}}\right)^{2} \left(\frac{D_{Observer-Moon}}{384,400 \text{ km}}\right)^{2}$. \par
In order to compare with the SP model, we used the procedure\footnote{\url{https://github.com/TKouyama/SP\_LunarCal\_example}} by \cite{Kouyama_2016}. We simulated Moon surface radiance for each pixel, assuming the images were taken by the ASTER instrument onboard Terra (EOS AM-1) satellite when the phase angle was 27.7$\degree$ $\pm$ 1$\degree$ in waxing phase moon from January 2000 to December 2015. However, for simplicity, we considered, in this work, an observer located at Terra altitude ($\sim$705 km), in the line of sight from Earth center to Moon center. To check the validity of this approximation, we compared the disk-integrated lunar irradiance considering the observation geometry parameters obtained by \cite{Kouyama_2019} and our parameters, for ASTER Moon observation on 14 April 2003. We showed that the irradiance error is typically below 0.2\% over the full wavelength range of the SP model, making the addition of the precise Terra orbital parameters negligible.\par
\begin{figure}
\centering
  \resizebox{\hsize}{!}{\includegraphics{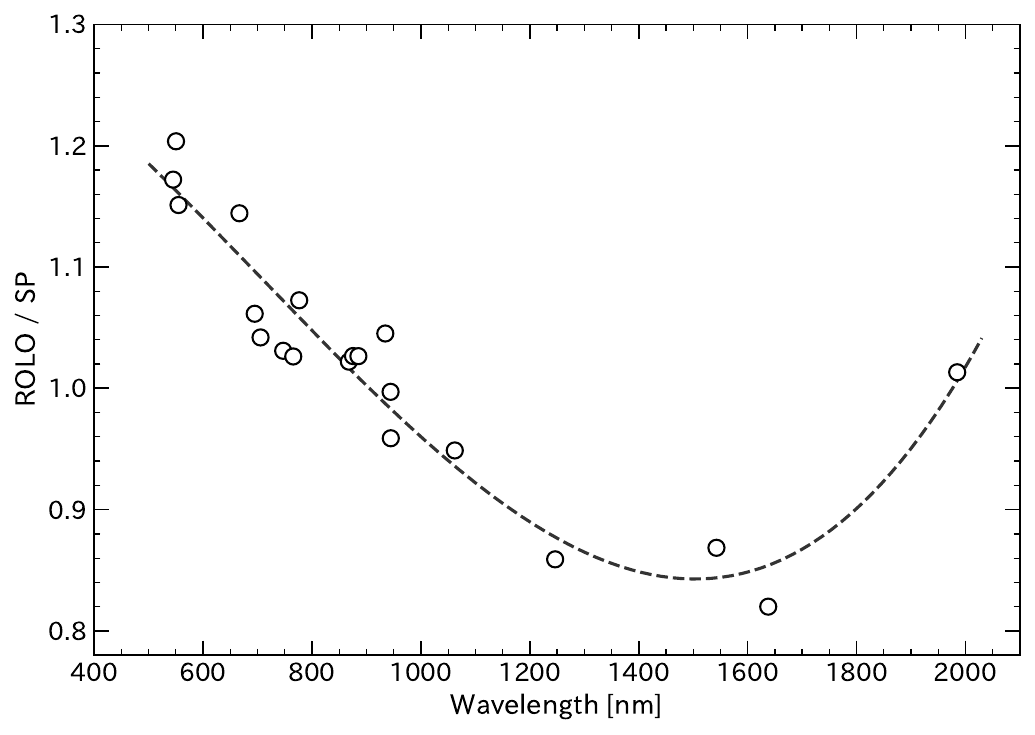}}
  \caption{Ratio of the irradiance from ROLO model and SP model. The dashed line represents the 3rd-degree polynomial $p(\lambda)$ best fit.}
  \label{fig:rolo_vs_SP}
\end{figure}

We found 78 geometrical configurations Sun-Terra-Moon with the defined phase angle. We computed the integrated irradiance for each of the time and averaged them to obtain a single spectrum of the Moon. We calculated the ROLO-band-integrated irradiance from the continuous SP model spectra considering Gaussian-shaped filter responsivity using nominal wavelength and filter width values from \cite{Kieffer_2005}. We then defined the correction factor as the best fit 3rd-degree polynomial function $p(\lambda)$ for the ratio of the ROLO irradiance over the SP irradiance (Fig. \ref{fig:rolo_vs_SP}). The ROLO correction factor fitting coefficients for the SP data extended up to 2053 nm are: $a_0 = 1.334$, $a_1 = -9.869 \times 10^{-5}$, $a_2 = -5.209 \times 10^{-7}$, and $a_3 = 2.455 \times 10^{-10}$.

\subsubsection{The M$^{3}$ model}
Because the SP model was derived for only wavelengths up to 2053 nm, the overlapping region with NIRS3 data is small (1800 -- 2053 nm) and does not allow checking whether the spectral slope of NIRS3 is also in agreement with other lunar observations. To verify if the assumption of a wavelength-independent scale factor is valid, we checked the consistency of our NIRS3 observations to Moon spectra projected in the same geometry as NIRS3 spectra obtained with instruments covering a larger portion of the wavelength range of NIRS3, that is the imaging spectrometer M$^3$ (400 -- 3000 nm) onboard Chandrayaan-1 \citep{Green_2011}. For this purpose, we used all data acquired during the OP2C1 mission phase of Chandrayaan-1. This choice was motivated by the wide spatial coverage of this data, and because this data were used to build the photometric model of the Moon used for M$^3$ data photometric correction \citep{Besse_2013}. In addition, because Chandrayaan-1 and M$^3$ suffer from thermal issues, which causes variations in obtained spectra during different orbits, the use of single mission phase dataset is more conservative. We used 175 L2 calibrated, thermally- and photometrically-corrected M$^3$ images to build a spectral photometrically-corrected radiance factor map ($i_0 = 30\degree, e_0 = 0\degree, \alpha_0 = 30\degree$) with wavelength ranges from 400 nm to 3000 nm, in a similar way as \cite{Yokota_2011} for SP data. To be consistent with the SP model, we created a latitude/longitude grid of $0.5\degree \times 0.5\degree$ for the 85 spectral channels. Because the M$^3$ images do not fully cover the lunar surface (about 82\% is covered), we considered a 2D spatial linear interpolation. We evaluated the uncertainties associated with this lack of data and with the linear interpolation, by computing the average M$^3$ spectra within the NIRS3 footprints for both the raw and interpolated spectral maps. Within the wavelength range of interest for this study (i.e., 1800 -- 3000 nm) the radiance factor varies between 0\% to 4\% (see Appendix \ref{appendix:interpolation_M3}). To derive the reflectance at the geometry ($i, e, \alpha$), we used the M$^3$ photometric model and the associated parameters obtained by \cite{Besse_2013}. It should be noted that these photometric parameters were derived exclusively for highlands and then optimized for this kind of materials which represents the majority of the lunar surface. Because M$^3$ were found to have similar spectral reflectance to ROLO -- slightly darker ($<$10\%) but with similar spectral slope \citep{Besse_2013_2, Ohtake_2013} -- we did not consider ROLO correction for these data. In fact, despite M$^3$ was not directly calibrated with ROLO observations, many comparisons have been made, showing that ROLO and M$^{3}$ are in agreement for all wavelengths and phase angles \citep{Buratti_2011, Hicks_2011, Green_2011}. We checked the wavelength-dependency of the average scale factor from M$^3$ data between 1850 nm and 2850 nm. We computed the ratio of the M$^{3}$ to the NIRS3 data for each wavelength channel. We observed that the necessary scale factor is almost wavelength-independent, with variations smaller than 5\% in the considered wavelength range (see Appendix \ref{appendix:correction_factor_SP_M3}). Because these variations are well below the uncertainties of the scale factor ($\sim 18\%$), we adopted a single wavelength-independent scale factor to correct the absolute reflectance level of NIRS3 data. This choice is further supported by the absence of lunar reflectance models beyond 2.95 $\mu m$, since the SP and M$^3$ models are respectively limited to 2.05 $\mu m$ and 2.95 $\mu m$, and by the known variability of the 2.8 $\mu m$ absorption band with surface region and local time (e.g., \citealt{Wohler_2017, Clark_2024}), which makes any wavelength-dependent correction poorly constrained at the longest wavelengths. \par

\subsubsection{Results of the absolute calibration}
The ROLO correction can now be applied to the SP spectra:
\begin{equation}
    r_{\text{norm}}(\lambda,i,e,\alpha) = r(\lambda,i,e,\alpha) \cdot p(\lambda),
\end{equation}
where $r(\lambda,i,e,\alpha)$ is the pre-ROLO-corrected SP reflectance. \newline
We can then directly compare the simulated spectra from SP and M$^{3}$ models projected to geometry of observation and illumination of NIRS3 spectra. By fitting the overlapping wavelength region between SP and NIRS3 (i.e., 1820 -- 2048 nm), and M$^3$ and NIRS3 (1850 -- 1976 nm) with a least-square algorithm, we derived scale factors for each of the 45 lunar NIRS3 spectra. The final scale factor $f_{\text{RCC}, j}$ for both SP and M$^3$ is given by the average of the 45 scale factors and the associated errors by the 1$\sigma$ value:
\begin{align}
    f_{\text{RCC,SP}} &= 1.04 \pm 0.09 \label{eq:scale_factors_sp} \\
    f_{\text{RCC,M$^{3}$}} &= 1.12 \pm 0.18 \label{eq:scale_factors_m3}
\end{align}
Detailed derivation of these scale factor is described in Appendix \ref{appendix:correction_factor_SP_M3}.\par
We finally considered that the necessary correction factor for the NIRS3 data can be represented by the mean of the two scale factors (Eqs. \ref{eq:scale_factors_sp} and \ref{eq:scale_factors_m3}), which give the following value of $\bar{f}_{\text{RCC}}$:
\begin{equation}
    \bar{f}_{\text{RCC}} = 1.08 \pm 0.10
\end{equation}
The absolute correction error was computed considering the propagated error of the two independent uncertainties of $f_{\text{RCC,SP}}$ and $f_{\text{RCC,M$^{3}$}}$. \par
Therefore, the Moon observations indicate that NIRS3 spectra should be upscaled by 8\% compared to the original calibration derived from ground-based measurements. The uncertainty associated with the absolute reflectance level is also improved compared to the original calibration, which was found to be $< 15\%$ \citep{Kitazato_2019}.

\subsection{Relative calibration of the data obtained during the different mission phases} \label{sect:relative_calib}

\begin{figure*}[ht!]
\plottwo{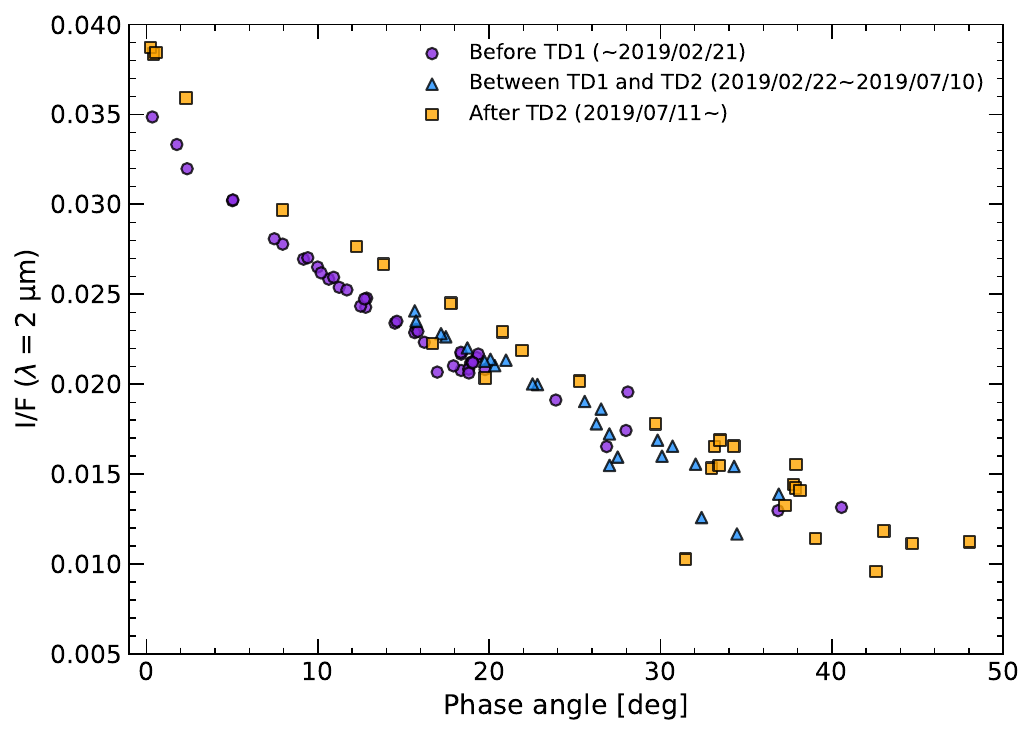}{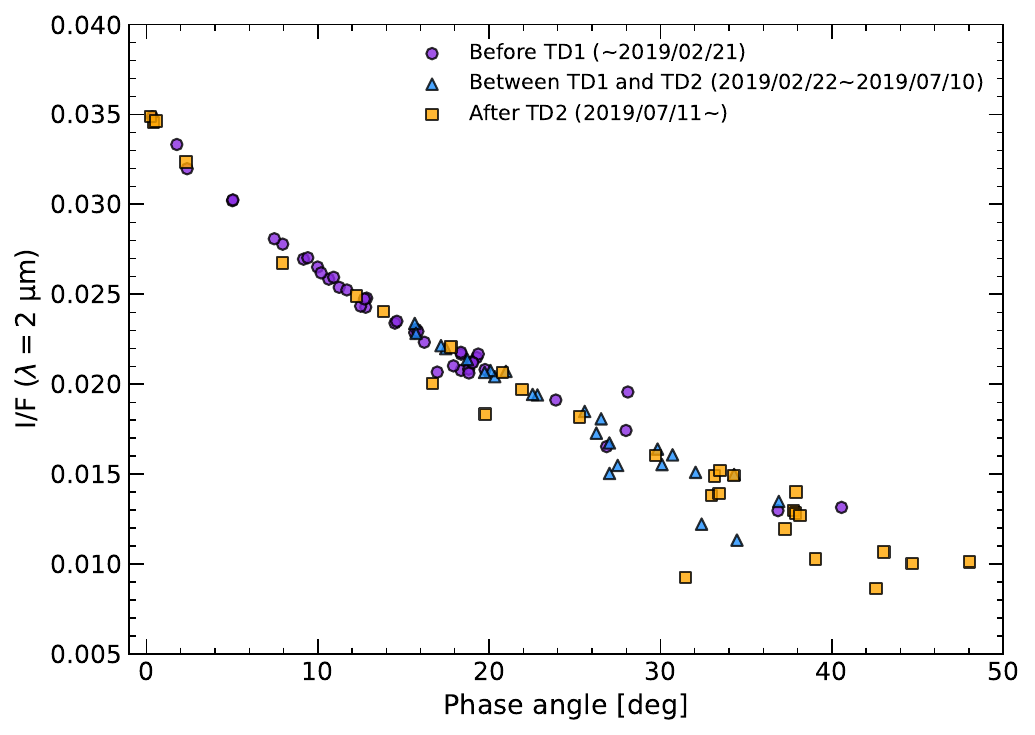}
\caption{Ryugu phase curves from NIRS3 data before TD1, between TD1 and TD2, and after TD2, at 2 $\mu m$ (left) before correction (right) after correction.
\label{fig:phase_curved_by_phases}}
\end{figure*}

We examined the data sorted by mission phases: before TD1 (2018/06/30 -- 2019/02/21), between TD1 and TD2 (2019/02/22 -- 2019/07/10), and after TD2 (2019/07/11 -- 2019/11/04). We observed that the radiance factor values were, for the entire Ryugu's surface, slightly higher for the data obtained between the two touchdowns compared to data acquired before TD1 and significantly higher for data obtained after TD2, clearly indicating the overcorrection of the RCC in the original calibration. Here, we considered that data acquired before TD1 were accurately calibrated thanks to the absolute calibration with the Moon (Sect. \ref{sect:moon_absolute_calib}). To quantitatively evaluate the overcorrection introduced by the calibration with the internal RAD calibration lamp, we decided to compare the phase curves corresponding to the different mission phases (Fig. \ref{fig:phase_curved_by_phases}). Phase curves were constructed by computing the mean radiance factor and mean phase angle for each observation sequence (i.e., one day of observations). The uncertainty associated with each data point represents the standard deviation of the radiance factor and phase angle values recorded within that sequence. Using a chi-squared ($\chi^2_{i} = \sum \frac{(f_{i} \cdot I/F_{i} - I/F_{TD1})}{f_{i}^{2}\sigma_i^{2} + \sigma_{TD1}^{2} }$, where $f_{i}$ is the scale factor, and with $i = \text{TD2}^{+}$ or $i = \text{TD1:TD2}$) minimization procedure, we estimated the best correction factor -- phase angle independent -- for the phase curves corresponding to the data after TD2 and the data obtained between TD1 and TD2. The uncertainties on the scale factors were computed by searching the range of values for which $\chi^{2}_{i} = \chi^{2}_{min,i} + 1$, corresponding to a 68\% confidence interval (i.e., 1$\sigma$ considering a Gaussian distribution). An example of correction at 2 $\mu m$ is shown in Fig. \ref{fig:phase_curved_by_phases}. The correction factor was found to be wavelength dependent (Appendix \ref{appendix:scale_factor_wavelength}, Fig. \ref{fig:scale_factors_vs_wavelength}). For data between TD1 and TD2 it started from 0.945 $\pm$ 0.015 at 1.8 $\mu m$ and continuously increases to 1.00 $\pm$ 0.02 at 3.1 $\mu m$. For data after TD2 the correction factor is decreasing with wavelength and ranges from 0.899 $\pm$ 0.003 at 1.8 $\mu m$ to 0.78 $\pm$ 0.02 at 3.1 $\mu m$. The correction factors drops or rises sharply in the wavelength range 3.1 -- 3.2 $\mu m$.  This evolution at the longer wavelength is likely due to the decrease of the NIRS3 signal-to-noise ratio in this region and/or to thermal tail removal residuals.\newline
At 2 $\mu m$, the scale factors $f_i$ are:
\begin{equation}
    \begin{cases}
    f_{\text{TD2}^{+}}(\lambda = 2 \mu m) = 0.897 \pm 0.004\\
    f_{\text{TD2:TD1}} (\lambda = 2 \mu m) = 0.946 \pm 0.018\\
\end{cases}
\end{equation}
This factor, derived for the 128 channels, allows to significantly improve the bias between the dataset obtained during the three mission phases. For instance, at 2 $\mu m$, the data after TD2 need to be decreased by $10.3 \pm 0.4\%$ to match the data obtained before TD1. Similarly, the data obtained between TD1 and TD2 should be downscale by $5.4 \pm 1.8\%$. These values mean that the data obtained after TD2 (resp., between TD1 and TD2) can be now used with data before TD1 with a precision $< 1\%$ (resp., $< 2\%$). This factor can be applied to the spectra which will be acquired during the extended mission phase, by multiplying the calibrated data by $f_{\text{TD2}^{+}}(\lambda)$. 

\subsection{Comparison with ONC-T observations}
\begin{figure}
\centering
  \resizebox{\hsize}{!}{\includegraphics{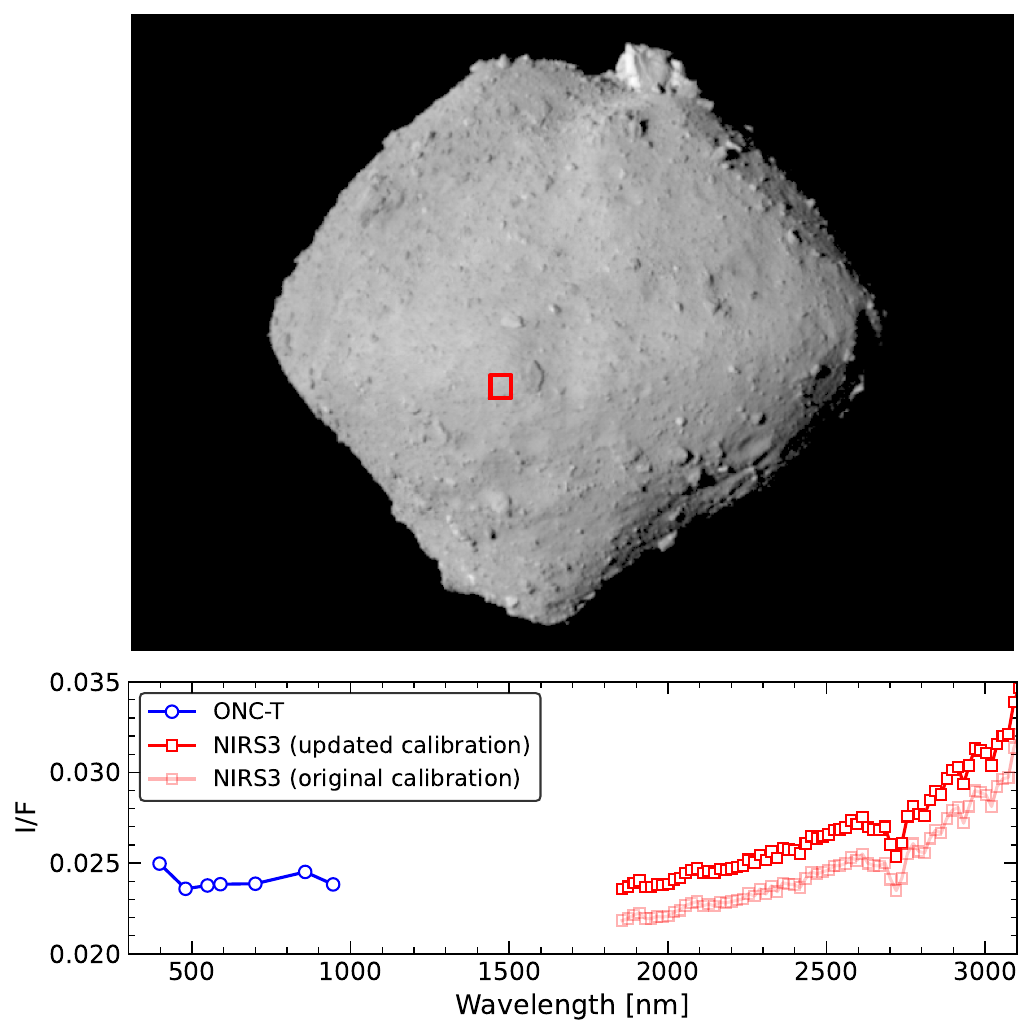}}
  \caption{(Top) ONC-T image acquired in \textit{ul} filter from home position ($\sim$20 km) on 2018-07-12T06:45:58.00 along with NIRS3 footprint (red square) on 2018-07-12T06:45:59.90. The image is cropped to highlight the position of the NIRS3 footprint. (Bottom) ONC-T radiance factor spectrum (blue circles) integrated from pixels inside the NIRS3 footprint, and NIRS3 radiance factor spectrum after applying calibration factors obtained in this work (red squares).}
  \label{fig:ONC_NIRS3_comp}
\end{figure}

It is important to check whether the updated calibrations give more consistent reflectance level with respect to ONC-T, than the originally calibrated data. This is due to the independent radiometric calibration of ONC-T using standard stars, achieving a better accuracy of 3\% \citep{Tatsumi_2019, Yumoto_2024}. If the calibration is accurate, the reflectance spectra can then be study from the near-UV and visible (390 -- 950 nm) to the near-infrared wavelength range (1800 -- 3200 nm). We computed the alignment between ONC-T and NIRS3 using the SPICE instrument and frame kernels. In ONC-T reference frame, NIRS3 footprints is located at pixels coordinates: \{[476.5, 454.3]; [492.7, 472.1]\}. These values are in perfect agreement with values derived by \cite{Tatsumi_2019}. We compared NIRS3 spectra with both ONC-T images of the Moon and Ryugu. For the Moon, we selected images acquired as close as possible to the NIRS3 acquisition time, with less than one-minute difference, for all ONC-T filters. For Ryugu, we considered ONC-T images taken with less than two seconds differences with respect to a NIRS3 spectrum. This duration was chosen to ensure that both instruments observe the same region of Ryugu's surface, thereby avoiding bias related to surface variations and observation conditions. We integrated the ONC-T radiance factor values for each pixel and inside the NIRS3 footprint (in ONC-T pixel coordinates). We compared the resulting spectrum with the NIRS3 spectrum (Fig. \ref{fig:ONC_NIRS3_comp}). \par 
As Ryugu exhibits a flat to slightly red spectrum from the visible to the near-infrared \citep{Moskovitz_2013, LeCorre_2018}, NIRS3 spectra should have a similar or slightly higher reflectance level than ONC-T. However, with the original calibration, NIRS3 spectrum was darker than ONC-T for this example acquired on 2018-07-12T06:45:59. After applying the updated calibration, NIRS3 spectrum exhibits a reflectance similar to ONC-T, confirming the improvement of the absolute reflectance level of NIRS3 data. While Fig. \ref{fig:ONC_NIRS3_comp} presents the results for a single combination of an ONC-T image and a NIRS3 spectrum, we also showed that this improvement of the reflectance level -- compared to the original calibration -- is observed for all NIRS3 spectra (Appendix \ref{appendix:NIRS3-ONC_comp}; Fig. \ref{fig:ONC_NIRS3_comp_all}). \par
\begin{figure}
\centering
  \resizebox{\hsize}{!}{\includegraphics{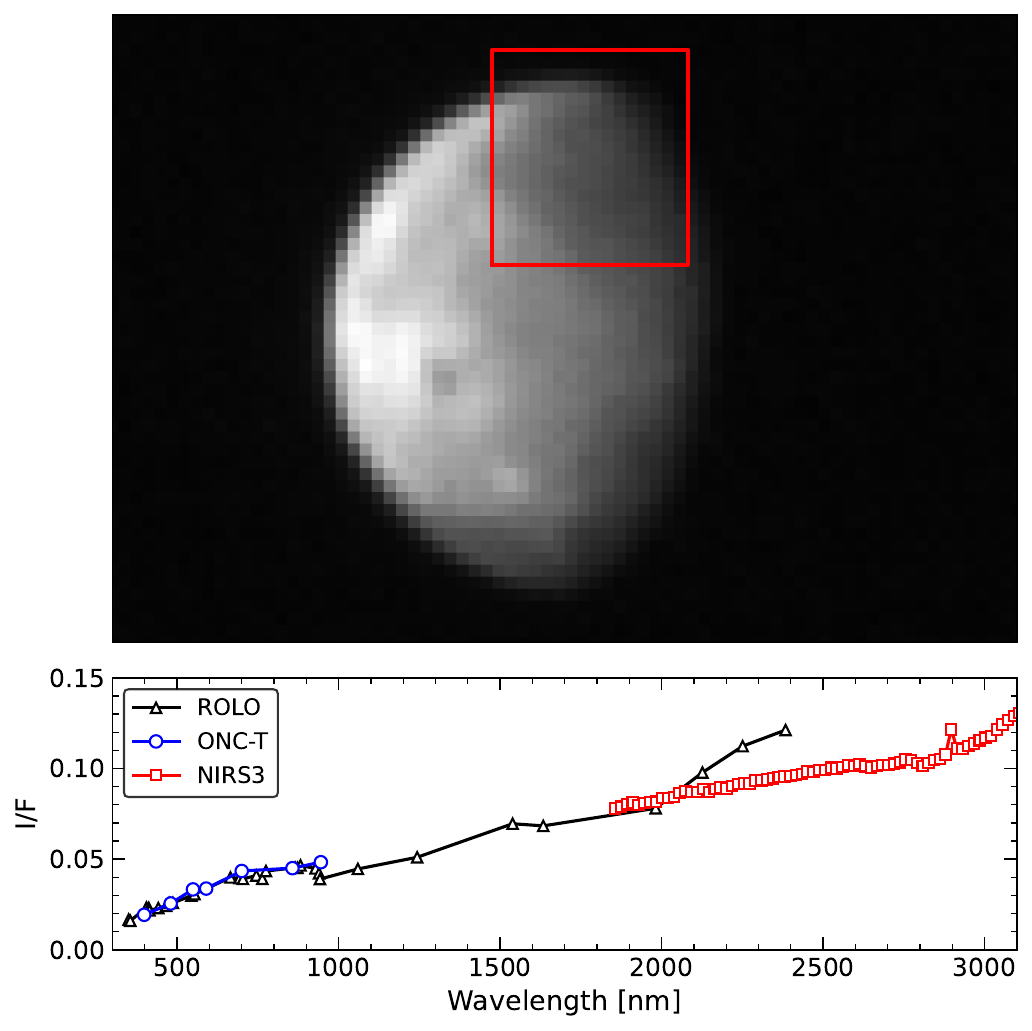}}
  \caption{(Top) ONC-T image of the Moon acquired in \textit{ul} filter on 2015-12-05T11:50:43.00 during the Earth swing-by, along with NIRS3 footprint (red square) on 2015-12-05T11:50:38.00. The image is cropped to highlight the position of the NIRS3 footprint. (Bottom) ONC-T radiance factor spectrum (blue circles) integrated from pixels inside the NIRS3 footprint, and NIRS3 radiance factor spectrum after applying calibration factors obtained in this work (red squares). The ROLO spectrum at the geometry of the Hayabusa2 instrument observations is also shown (black triangle).}
  \label{fig:ONC_NIRS3_moon_comp}
\end{figure}

The results for the Moon shows also a very good agreement between ONC-T and NIRS3 reflectance level, as well as with the ROLO model. Fig. \ref{fig:ONC_NIRS3_moon_comp} shows the comparison for a simultaneous lunar observation of ONC-T and NIRS3. Only one lunar observation was obtained with a similar acquisition time between the two instruments, which is the sequence starting at 2015-12-05T12:48:10 with the ONC-T \textit{v} filter. We compared the data from ONC-T and NIRS3 with the ROLO model, which is defined from 350 to 2450 nm with 32 wavelength channels. Geometry information for the ROLO model inputs were computed from Hayabusa2 SPICE kernels at the acquisition time of the ONC-T \textit{ul} filter image (see also Table \ref{tab:NIRS3_lunar_observations_properties}). The ONC-T and NIRS3 spectra exhibits an expected red featureless spectrum. ONC-T is in very good agreement in both spectral slope and reflectance level to the ROLO model spectrum. The wavelength range gap between ONC-T and NIRS3 appears to be well-matched with the ROLO model. NIRS3 is congruent with respect to ROLO, except for the two last ROLO filters, which have also higher reflectance compared to the filters at shorter wavelength \citep{Kieffer_2005}. Therefore, the updated calibration of NIRS3 proposed in this work is consistent with ONC-T spectral slope and reflectance level, improving the original calibration of the instrument. \par

\subsection{Conclusion on updated NIRS3 calibration}
Future studies using NIRS3 data may then use this absolute correction factor. The correction factor $\bar{f}_{RCC}$ should be simply multiplied with the reflectance data. The updated data (L2D and higher levels) for Ryugu observations will be archived in PDS and DARTS. This scale factor should be applied to data obtained at any mission phases from approach to extended mission and may be then applied for the future NIRS3 observations of (98943) Torifune and 1998 KY26. For the extended mission NIRS3 data, the wavelength-dependent $f_{\text{TD2}^+}$ scale factor should be also considered, which means that all data must be multiplied by $f_{\text{TD2}^+} \times \bar{f}_{\text{RCC}}$. \par
As all correction factors are phase angle-independent, the updated calibration do not modify the shape of the phase curve derived in previous photometric work using NIRS3 data \citep{Domingue_2021, Pilorget_2021, Longobardo_2022}. Only the absolute values are modified. \par

\section{Disk-resolved global near-infrared photometric properties of Ryugu} \label{sec:disk_resolved}
The updated calibration of NIRS3 opens new opportunities to perform near-infrared spectroscopic and photometric studies of Ryugu, especially because using all data from all mission phases together is now possible. In this work, we were interested in deriving new photometric properties of Ryugu, with the final goal of providing photometrically corrected NIRS3 reflectance data. 

\subsection{Dataset and data analysis}
For this photometric analysis, we used thermal emission removed I/F (L2D) data acquired during the Hayabusa2 proximity phase from June 2018 to November 2019 with our updated calibration explained previously. This dataset covers a phase angle range from 0.13$\degree$ to 48.2$\degree$. More details about observation conditions are given in Appendix \ref{appendix:obs_conditions}. Observations acquired at extreme illumination or emission geometries ($i,e > 70\degree$) were excluded due to their low signal-to-noise ratio and to mitigate strong shadowing and limb effects. Observations obtained at high instrument temperatures ($T > 350K$) and phase angle larger than 15$\degree$ were likewise discarded, as they were found to significantly alter the derived spectral slope values. Finally, observations with a spatial resolution worse than 40 meters  -- inferior to that of the home position observations -- were also rejected. After applying these filters, a total of 249\,639 spectra were retained for this study, out of the 407\,150 spectra acquired during the proximity phase. Because many observations were obtained with a phase angle close to 20$\degree$, a binning is necessary to mitigate this bias in the data distribution. For this, we performed a 3D averaging binning ($i,e,\alpha$) following the method described in \cite{Yokota_2021}, that is considering the following binning width: $\Delta i = 1\degree$, $\Delta e = 1\degree$, and an adaptative phase angle width: 

\begin{eqnarray}
        \Delta \alpha =
    \begin{cases}
    0.1\degree, & 0\degree \le \alpha < 2\degree,\\
    0.2\degree, & 2\degree \le \alpha < 5\degree,\\
    1\degree, & 5\degree \le \alpha \le 50\degree.
    \end{cases}
\end{eqnarray}

\subsection{Choice of photometric model} \label{sec:choice_model}
Many photometric models have been proposed in the literature. We therefore tried several models to investigate the one that provides the best photometric correction for our data. We selected five different photometric models to test: (1) Hapke \citep{Hapke_2012}; (2) Linear-magnitude phase function associated with Lommel-Seeliger disk function (hereafter designated as "LinMagLS"); (3) Linear-magnitude phase function combined with Akimov disk function (hereafter designated as "LinMagAkimov"); (4) Exponential phase function associated with McEwen disk function with exponential polynomial partition function (hereafter designated as "ExpMcEwen"); and (5) ROLO phase function combined with Lommel-Seeliger disk function (hereafter designated as "ROLO-LS"). The models have free parameters ranging from 2 (for LinMagLS and LinMagAkimov) to 7 (for ExpMcEwen and ROLO-LS). These models were chosen because they were used successfully for Ryugu and/or Bennu photometric studies \citep{Tatsumi_2020, Golish_2021_1, Golish_2021_2, Zou_2021}. The models are described in more details in Appendix \ref{appendix: models}. \par
To compare the fitting from the different models, we computed the root-mean-square (RMS) for each wavelength channel. The results at 2.0 $\mu m$ are given in Table \ref{tab:RMS_comp_models}. RMS comparison between the models is shown in Appendix \ref{appendix:rms} (Fig. \ref{fig:RMS_diff_with_wvl}). Because of the rough surface of Ryugu, the observations are affected by an opposition surge, which is not modeled in LinMagLS and LinMagAkimov. Therefore, even if the fit with these two models works well for the majority of the phase angle range, they do not fit the data accurately at phase angle smaller than 5$\degree$. Over the full wavelength range, Hapke provides a better fit than any other models. ExpMcEwen and LinMagAkimov have a similar RMS wavelength dependency compared to Hapke with values only 0.7 to 4.0\% higher than Hapke. The ROLO-LS and the LinMagLS models have an RMS of about 17 to 20\% higher than the one obtained with the Hapke model. \par
Considering these results, we finally chose to use the Hapke model for the photometric correction of the NIRS3 data of Ryugu.

\begin{table}[]
    \centering
    \caption{RMS for the different models for Ryugu NIRS3 data. }
    \begin{tabular}{cc}
    \hline
    \hline
      Model & RMS at 2.0 $\mu m$ ($\times 10^{-3}$)\\
    \hline
       Hapke & 1.45\\
       LinMagLS & 1.72\\
       LinMagAkimov & 1.50\\
       ExpMcEwen & 1.48\\
       ROLO-LS & 1.70\\
    \hline
    \end{tabular}
    \label{tab:RMS_comp_models}
\end{table}

\subsection{Hapke modeling: methods and results} \label{sect:methods_hapke}
The fitting with the Hapke model was carried out following a four-steps procedure. This method was chosen to disentangle the five highly-degenerated Hapke parameters. The fits were performed assuming all wavelength channels are independent with each other, therefore with no initial guess dependency with previous channels. The first step consists in fixing the $\omega$, $g$, and $\bar{\theta}$ parameters to values determined in the visible wavelength range from ONC-T data by \cite{Tatsumi_2020}; leave $B_{sh,0}$ and $h_{sh}$ free within respectively \{[0.0, 2.0]\} and \{[0.0, 0.15]\} bounds, and fitting only the data acquired at low solar phase angle ($< 20\degree$). In the second step, we fixed $B_{sh,0}$ and $h_{sh}$ to the values previously determined, as well as $\bar{\theta}$ from ONC-T, in order to constrain the parameters $\omega$ and $g$. We fitted using the full dataset, and constraining $\omega$ and $g$ within \{[0.0, 0.2]\} and \{[-0.4, 0.1]\} bounds, respectively. For the third run of the fitting procedure, we fixed $\omega$, $g$, $B_{sh,0}$, and $h_{sh}$ and determined $\bar{\theta}$ within the bounds \{[10, 35]\}. The last step was used to determine reliable uncertainties for each of the parameters. It consists of leaving all parameters free with initial values set to the values determined from the three previous runs with a given relative step size. With this procedure $\bar{\theta}$ is strongly constrained by the fixed value of the other parameters. We conducted a sensitivity analysis to estimate how the spectra of the parameters are modified when $\bar{\theta}$ is fixed to various values. The analysis is shown in Appendix \ref{appendix:sensitivity}. We show that this does not impact the spectral behavior of the Hapke parameters and hence of the NIRS3 spectra, but only the absolute values by less than 5\%.\par

\begin{figure*}[ht!]
\plottwo{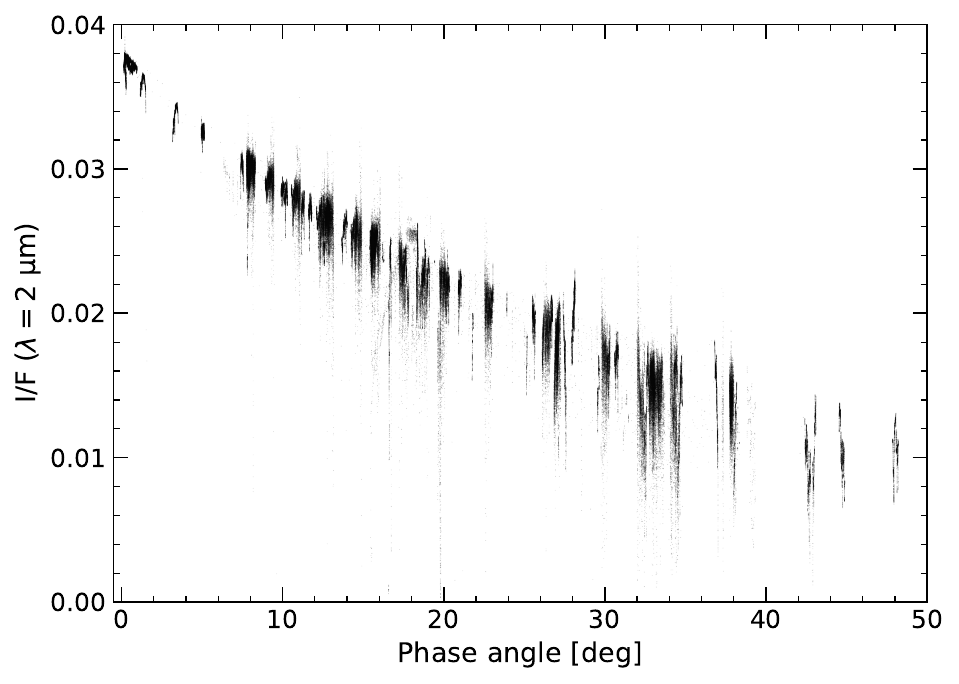}{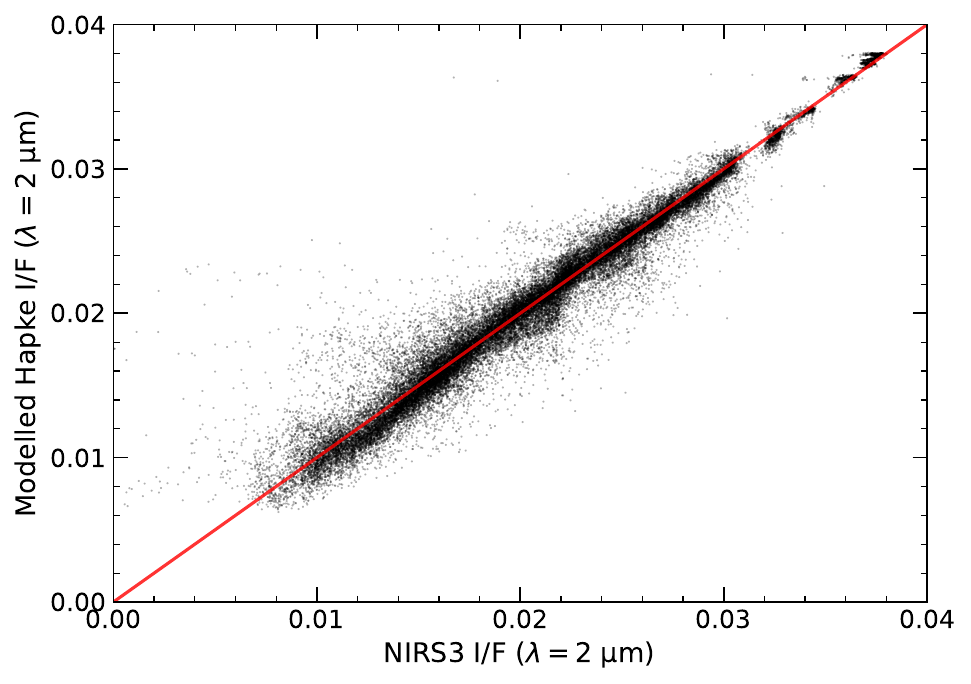}
\caption{(Left) Ryugu phase curve from NIRS3 data acquired during the proximity phase at 2 $\mu m$. (Right) Modelled vs NIRS3 radiance factor values at 2 $\mu m$. The red solid line represents the $y = x$ straight line.
\label{fig:resolved_Hapke-2THG_phase_curve}}
\end{figure*}

\begin{table*}[]
    \centering
    \caption{Hapke parameters from Ryugu global disk-resolved analysis.}
    \begingroup
    \renewcommand{\arraystretch}{1.5}
    \begin{tabular}{ccccc}
    \hline
    \hline
        & This work & 
        \cite{Tatsumi_2020} & \cite{Domingue_2021} & \cite{Pilorget_2021}\\
        Instrument, Wavelength & NIRS3, 2 $\mu m$ & 
        ONC-T, 0.95 $\mu m$ & NIRS3, 2 $\mu m$ & NIRS3, 1.89 $\mu m$\\
        Hapke parameters & & & & \\
       \hline
       $\omega$ & 0.037 $\pm$ 0.001 & 0.046 $\pm$ 0.008 & 0.049 $\pm$ 0.001 & 0.051 $\pm$ 0.010\\
       $g$ & -0.340 $\pm$ 0.005 & -0.377 $\pm$ 0.001 & -0.318$^{+0.007}_{-0.004}$ & -0.30$^{+0.05}_{-0.11}$ \\
       $B_{sh,0}$ & 1.06 $\pm$ 0.03 & 0.98 $\pm$ 0.02 & 0.99 $\pm$ 0.02 & 0.94$^{+0.02}_{-0.01}$ \\
       $h_{sh}$ & 0.105 $\pm$ 0.005 & 0.075 $\pm$ 0.008 & 0.110 $\pm$ 0.003 & 0.10$^{+0.03}_{-0.05}$ \\
       $\bar{\theta}$ [deg] & 29 $\pm$ 2 & 28 $\pm$ 6 & 29$^{+3}_{-2}$ & 32$^{+13}_{-22}$ \\
       \hline
       RMS & $1.45 \times 10^{-3}$ & -- & -- & -- \\
    \hline
    \end{tabular}
    \endgroup
    \label{tab:hapke_params_global}
\end{table*}

\begin{figure}
\centering
  \resizebox{\hsize}{!}{\includegraphics{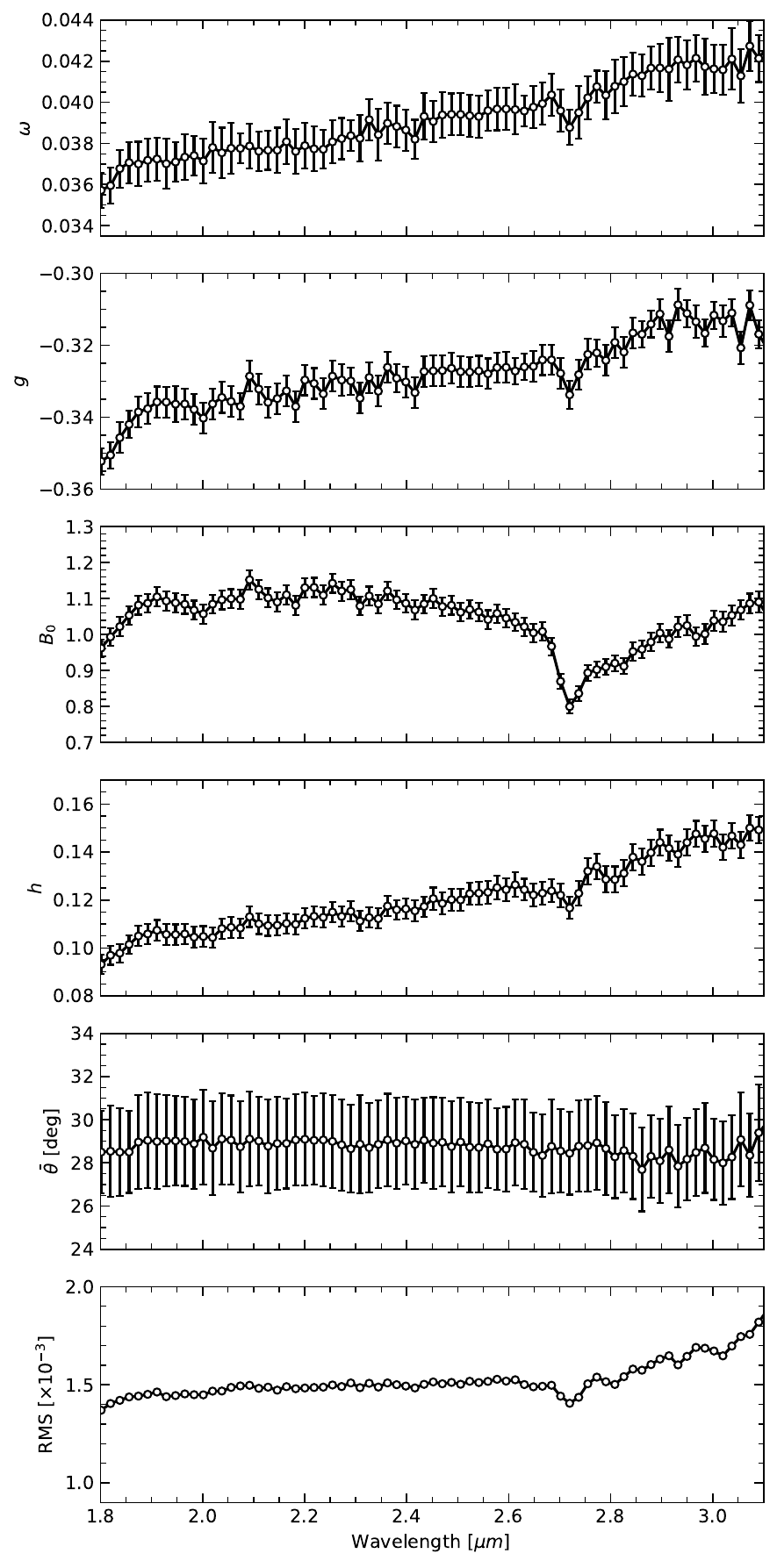}}
  \caption{Spectral evolution of the Hapke parameters for Ryugu, in the near-infrared  (1.8 -- 3.1 $\mu m$).}
  \label{fig:hapke_params_with_wvl}
\end{figure}

The results at 2 $\mu m$ are given in Table \ref{tab:hapke_params_global} and Fig. \ref{fig:resolved_Hapke-2THG_phase_curve}. The wavelength dependency of the parameters is shown in Fig. \ref{fig:hapke_params_with_wvl}. Hapke model provides a very good fit of the data for all wavelength and for the entire phase angle range, including the opposition effect (Fig. \ref{fig:resolved_Hapke-2THG_phase_curve}). Our derived Hapke parameters are consistent with previous estimation from Ryugu disk-resolved analysis of NIRS3 data by \cite{Domingue_2021} and \cite{Pilorget_2021}, but also similar to the parameters derived by \cite{Tatsumi_2020} from ONC-T data. Within uncertainties, $g$, $h_{sh}$, and $\bar{\theta}$ exhibits similar values to one or both previous NIRS3 photometric studies. On the other hand, $\omega$ and $B_{sh,0}$ are slightly different. The single-scattering albedo derived from this work is smaller ($0.037 \pm 0.001$) than the previous NIRS3 estimation ($\sim0.050$) but also to the visible wavelength range parameter. This different value is solely due to the different version of the Hapke model used, because we took into account here the porosity correction factor which mainly acts as a multiplicative factor with the single-scattering albedo. The porosity correction factor also appears in the multiple scattering function, however, the contribution of this term is very weak due to the dark surface of Ryugu. For the SHOE amplitude parameter ($B_{sh,0}$), our value is marginally higher ($1.06 \pm 0.03$) than the previous works (0.94 -- 0.99). This minimal difference may be due to the better coverage of the opposition effect thanks to the data obtained after TD2 compared to the previous NIRS3 photometric studies. Regarding variations of the Hapke parameters with wavelength (Fig. \ref{fig:hapke_params_with_wvl}), we observed that the single-scattering albedo has an expected behavior with a red spectrum and a weak 2.7 $\mu m$ absorption feature visible. The $g$ and $h_{sh}$ parameters follow the same trend (i.e, increasing values with increasing wavelength, but a small decrease associated with the 2.7 $\mu m$ feature). $B_{sh,0}$ is roughly constant between 1.8 and 2.6 $\mu m$, with these values ranging from 1.02 to 1.16 in this wavelength domain. The value drops sharply around 2.7 $\mu m$ reaching as low as 0.8. This correlation between the SHOE amplitude and the single-scattering albedo has been discussed in \cite{Domingue_2021} and is physically explained by the definition of the opposition effect amplitude (ratio of the light initially scattered to the total scattering in the opposition direction). Because the albedo decreases in the absorption feature, the incoming photons are more absorbed during the first scattering, leading to a decrease in the opposition amplitude. However, the fact that the shape of the feature is not exactly similar between $B_{sh,0}$ and $\omega$ is attributed by \cite{Domingue_2021} to the complex physical properties of the Ryugu surface, and in particular to grain size and/or surface roughness. Finally, the roughness ($\bar{\theta}$) parameter was found to be constant within error bars ($\pm 2\degree$) in the NIRS3 wavelength range with values between 25.3$\degree$ and 27.4$\degree$. This behavior may be the results of the fitting process. Because of the lack of constraints at large phase angles and the relatively large number of parameters in the Hapke model, we had to fit $\bar{\theta}$ at the penultimate step of the procedure with all other parameters fixed, resulting in a limited degree of freedom for this parameter.

\subsection{Phase reddening on Ryugu} \label{sect:phase_reddening}
Phase reddening -- which is defined as the increase of the spectral slope with increasing solar phase angle -- has been observed on the Ryugu surface by previous studies \citep{Tatsumi_2020, Longobardo_2022}. From ONC-T in the visible wavelength range, \cite{Tatsumi_2020} show that Ryugu exhibits a weak phase reddening of $(2.0 \pm 0.7) \times 10^{-3} \mu$m$^{-1}$ deg$^{-1}$. \cite{Longobardo_2022} use a limited NIRS3 dataset, with phase angle ranging from 15$\degree$ to 40$\degree$, to derive phase reddening properties of Ryugu in the near-infrared wavelength range. They found a phase reddening of $(2.1 \pm 0.3) \times 10^{-3} \mu$m$^{-1}$ deg$^{-1}$ in agreement with the visible phase reddening. We extended this near-infrared phase reddening analysis using all proximity phase observations, and hence covering a wider solar phase angle from opposition to 50$\degree$. \par
We fitted a straight line to the spectral slope values -- computed between 2.1 and 2.5 $\mu$m -- as a function of the phase angle (Appendix \ref{appendix:phase_reddening_resolution}). The slope coefficient corresponds to the phase reddening value. We estimated the uncertainties from the square root of the diagonal terms of the covariance matrix. We obtained a phase reddening of $(1.1 \pm 0.2) \times 10^{-3} \mu$m$^{-1}$ deg$^{-1}$. This value is lower than the previous estimation by \cite{Longobardo_2022}. It indicates that Ryugu exhibits a very weak phase reddening, comparable to the objects that presented the lowest observed phase reddening in the near-infrared such as Bennu ($(-0.04 \pm 0.06) \times 10^{-3} \mu$m$^{-1}$ deg$^{-1}$, \citealt{Fornasier_2020}) and Eros ($\sim 1.0 \times 10^{-3} \mu$m$^{-1}$ deg$^{-1}$, \citealt{Clark_2002}). \par
Ryugu exhibits an unusual phase reddening behavior relative to other small bodies. \cite{Longobardo_2022} showed that the near-infrared phase reddening of small bodies is typically three to four times lower than the visible phase reddening, whereas Ryugu displayed comparable values in both wavelength ranges. After applying the updated calibration, we find that the Ryugu’s near-infrared phase reddening is indeed lower than its visible phase reddening, though the difference remains smaller than for other asteroids. This distinct behavior between Ryugu and Bennu is therefore more likely related to variations in physical properties rather than composition. This interpretation is supported by the returned samples, which have shown that Bennu and Ryugu share a very similar red slope from the visible to the near-infrared (e.g., \citealt{Fukai_2026}), indicating similar materials and optical properties between the two bodies, properties that alone cannot account for their different phase reddening behavior. Phase reddening is generally associated with surface micro-roughness, which affects the multiple scattering properties of the surface, with the multiple scattering contribution increasing with wavelength among fine grains \citep{Schroder_2014}. Ryugu’s exposed boulders may then retain a higher fraction of micron-sized dust or impact-generated fine regolith, whereas Bennu’s boulders, while highly porous, may lack such fine particulate coatings, which would explain Bennu’s near-zero phase reddening. This difference in fine-particle abundance was already proposed to explain the visible spectral properties divergence between Ryugu and Bennu \citep{Yumoto_2024b}, and our near-infrared photometric results are consistent with that interpretation. Such an accumulation of fine grains on Ryugu could be explained by its higher surface gravity relative to Bennu (K. Yumoto et al., \textit{under review}).

\subsection{Regional analysis}

\begin{figure*}
\centering
  \resizebox{\hsize}{!}{\includegraphics{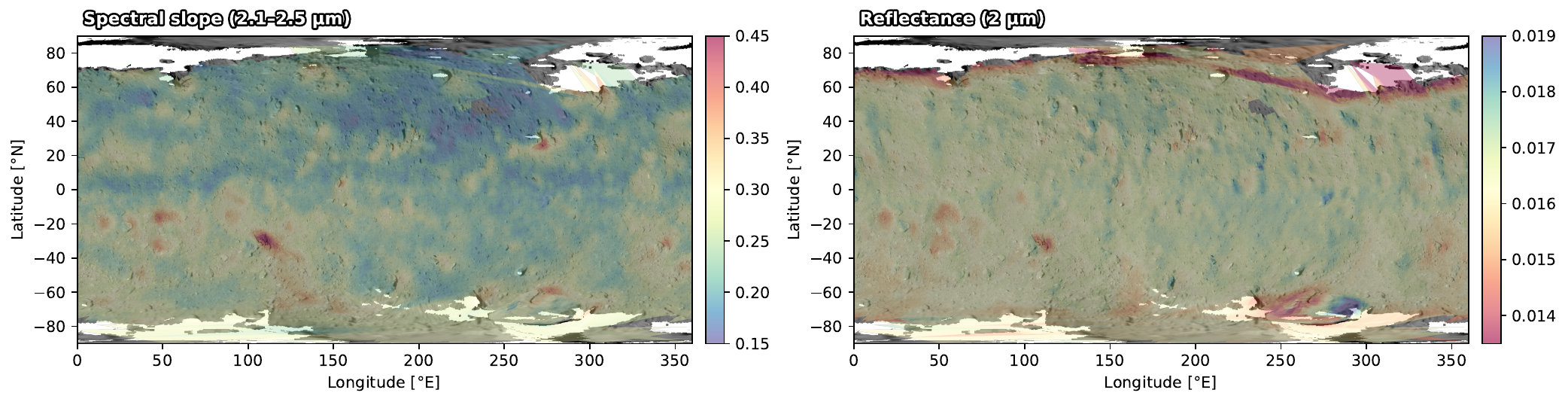}}
  \caption{Map of the (left) spectral slope computed from 2.1 and 2.5 $\mu m$ in $\mu m ^{-1}$; (right) reflectance at 2 $\mu  m$. The two maps were created using photometrically corrected reflectance data, obtained on 11 July and 19 July 2018. The footprints were rasterized into a $0.1\degree \times 0.1\degree$ latitude and longitude grid and the median value was computed for each grid element.}
  \label{fig:refl_slope_map}
\end{figure*}

We investigated potential regional variations of the photometric properties with this dataset by creating photometric parameter maps. To perform this regional analysis, we first defined a $10\degree \times 10\degree$ latitudinal-longitudinal grid on the surface of Ryugu. We limited the regions from -30$\degree$N to 20$\degree$N because of the spatial coverage of NIRS3. We attributed to each grid element the NIRS3 spectra based on the footprint positions. If a footprint overlaps several grid elements; we attributed the spectra to these different grid elements. We then apply a similar procedure to the global analysis for each grid element, removing extreme geometries, high instrument temperatures, and low spatial resolution observations, and performing 3D averaging binning. \par
After the binning procedure, we computed the photometric quantities of interest for each grid element: phase ratio, phase reddening, and photometric model parameters. To assess opposition effect, phase ratio is usually computed with reflectance values at 0.3$\degree$ and 5$\degree$ \citep{Belskaya_2000}. However, because of the lack of data at this very small phase angle, the phase ratio might be highly model-dependent. Therefore, we computed the phase ratio between 5$\degree$ and 15$\degree$, which provides insights into the SHOE properties. The phase reddening was computed following the method described in Sect. \ref{sect:phase_reddening}. We also derived parameter maps from the Hapke model, using the four steps procedure described in Sect. \ref{sect:methods_hapke}. These maps are presented in Appendix \ref{appendix:maps} (Fig. \ref{fig:hapke_maps}). \par
\begin{figure*}
\centering
  \resizebox{0.7\hsize}{!}{\includegraphics{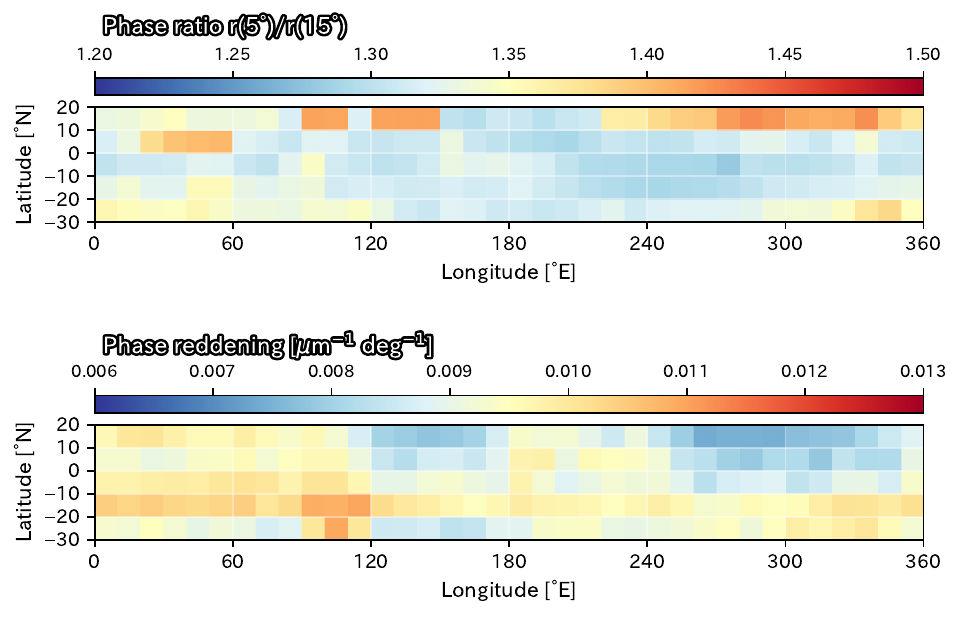}}
  \caption{Map of the (top) phase ratio $I(5\degree)/I(15\degree)$ at 2.0 $\mu m$ and (bottom) phase reddening (with spectral slope computed between 2.1 $\mu m$ and 2.5 $\mu m$). The maps were computed using a $10\degree \times 10\degree$ grid at the Ryugu surface. The particularly high phase ratio and low phase reddening values for some grid elements located at latitude above 10$\degree$N are due to a narrower phase angle range (see Appendix \ref{appendix:obs_conditions}, Fig. \ref{fig:phase_minmax}). All grid elements above this latitude were discarded for the computation of the average east and west properties.}
  \label{fig:ratio_reddening_maps}
\end{figure*}

Fig. \ref{fig:ratio_reddening_maps} shows the phase ratio and phase reddening maps. Because NIRS3 mainly observed Ryugu around its equatorial region, higher latitude regions have a poor number of observations and a narrow phase angle range (see Appendix \ref{appendix:obs_conditions}, Fig. \ref{fig:phase_minmax}). We therefore restrict our analysis to the reliable equatorial region, within -20$\degree$N to 10$\degree$N. In this region, both the phase ratio and phase reddening exhibit a weak dichotomy between the western and eastern hemispheres. The phase ratio of the western hemisphere (150--300$\degree$E; \citealt{Cho_2021}) is $1.305 \pm 0.002$ and $1.323 \pm 0.003$ for the eastern hemisphere (300--150$\degree$E). The average phase reddening of the eastern hemisphere is $0.0056 \pm 0.0001$ $\mu m^{-1}$deg$^{-1}$, and the average value for the western hemisphere is $0.0052 \pm 0.0001$ $\mu m^{-1}$deg$^{-1}$. Both quantities are therefore higher in the eastern hemisphere. \par
This east-west dichotomy has been already reported by various studies in both visible and infrared spectroscopic studies \citep{Sugita_2019, Barucci_2019, Tatsumi_2020}. However, because opposition observations by \cite{Yokota_2021_opposition} did not observe such hemispherical variations, \cite{Yokota_2022} hypothesize that these differences may come from variations in physical properties rather than from albedo alone. Their preliminary regional photometric study with ONC-T data indeed shows east-west variations of the Hapke roughness parameter $\bar{\theta}$ (Fig. \ref{fig:hapke_maps}). Our phase ratio and phase reddening results confirms this hypothesis: the slightly stronger opposition effect and stronger phase reddening for the eastern hemisphere, both indicating higher surface roughness, are visible with the NIRS3 observations. \par
We suggest that this east-west dichotomy of the near-infrared optical properties reflects differences in distribution of boulders and the evolution history of Ryugu. The western bulge exhibits fewer boulders than the eastern hemisphere \citep{Sugita_2019, Michikami_2019} and the crater-size frequency distribution shows that it formed after the resurfacing of the eastern hemisphere \citep{Cho_2021} possibly from rotationally induced deformations by the high spinning rate \citep{Hirabayashi_2019}, indicating that the western hemisphere corresponds to a younger terrain. This difference in surface age is confirmed by the variations of photometric properties. The higher phase reddening and phase ratio in the eastern hemisphere, together with the retrieved Hapke parameters -- a narrower $h_{sh}$ (and correspondingly a lower porosity factor $K$) and a higher $B_{sh,0}$ -- are consistent with higher roughness and micro-porosity associated with regolith accumulation in that hemisphere. The higher $B_{sh,0}$ in particular may reflect both this higher roughness/porosity and possibly a higher proportion of space-weathered, opaque grains rich in nanophase iron particles. Based on the thermal inertia values derived by \cite{Shimaki_2020}, we also showed that the eastern hemisphere may exhibit a slightly lower thermal inertia ($235.6 \pm 0.5$ tiu) than the western bulge ($241.7 \pm 0.5$ tiu), consistent with a higher amount of fine particulates in this region. \par
This regolith accumulation may be linked to the difference in surface age, implying different impact and thermal histories for the two hemispheres. The high porosity of the Ryugu's boulders \citep{Okada_2020} damps thermal stress and favors compaction over fragmentation upon impact \citep{Cambioni_2021}; becoming susceptible to thermal fatigue and beginning producing fine regolith only after sufficient exposure time. Given that the western bulge was resurfaced 2-9 Myr ago \citep{Cho_2021}, it has not had enough time to develop an equivalent abundance of fine regolith, whereas the older eastern hemisphere has experienced the longer impact history needed to accumulate the porous regolith reflected in its photometric properties.

\section{Conclusions}
We provided an updated in-flight calibration of the NIRS3 near-infrared spectrometer on board Hayabusa2. The purpose of this calibration was to improve two important aspects: (i) the reflectance level of NIRS3 was lower than expected compared to the ONC-T multispectral camera; (ii) due to touchdown operations, calibration using the internal lamp was not accurate enough to allow the use of data obtained after TD2. We used NIRS3 spectra of the Moon acquired before the arrival at Ryugu to perform the absolute calibration of the instrument. By comparing with various lunar photometric models -- namely ROLO, SP, and M$^3$ -- we derived correction factors to be applied to the data with the original calibration. To solve the calibration issues observed after touchdowns, we compared the phase curve obtained at the different mission phases, showing that the necessary correction factors are phase angle independent, but slightly dependent on the wavelength. \par
We found that the data obtained between TD1 and TD2 should be slightly downscale with typical values within the NIRS3 wavelength range between 0 and 6\% (with an uncertainty of about 1.8\%). On the other hand, data obtained after TD2 are significantly brighter and redder than the data obtained before TD1. These data should be decreased by 10\% at 1.8 $\mu$m and 21\% at 3.1 $\mu$m. After applying these correction factors, the updated calibration allows the use of the full dataset of NIRS3 observations acquired during the entire proximity phase, which is particularly valuable as the data after TD2 contains data covering different phase angle ranges than the limited dataset up to TD1. From the Moon observations, we showed that all NIRS3 data should have a reflectance level 8\% brighter than with the original calibration. This new calibration also significantly improves the possibility to perform studies combining ONC-T and NIRS3 observations. \par
From the updated calibration, we used the full NIRS3 proximity phase dataset to compute new photometric properties and parameters. We found similar values within uncertainties for the photometric parameters compared to previous studies. However, we also noticed a smaller phase reddening compared to the previous estimation, but still slightly higher than the one on Bennu, indicating possible variations in roughness/grain size distribution between the two bodies. From regional analysis, we found an east-west dichotomy of near-infrared phase ratio and, for the first time, of phase reddening. This dichotomy was already reported by visible albedo and slopes, and investigated by geomorphological studies. The variations of the near-infrared photometric quantities are consistent with a more important roughness and a higher abundance of fine-grained materials in the eastern hemisphere compared to the western bulge. The derived global photometric parameters are used to produce NIRS3 L3A data (photometrically-corrected reflectance spectra) which will be made available on the NASA PDS and JAXA DARTS.

\section*{Data availability}
NIRS3 I/F data (L2C) used in this work are available through JAXA DARTS and NASA PDS Small Bodies Node \citep{https://doi.org/10.17597/isas.darts/hyb2-00400}. NIRS3 thermal excess removed I/F data (L2D) and footprint geometry are available through JAXA DARTS (\citealt{Ichikawa_2026a} and \citealt{Ichikawa_2026b}). Radiance factor co-registered ONC-T data (level 2drc) are available through JAXA DARTS and NASA PDS \citep{https://doi.org/10.17597/isas.darts/hyb2-00200}. Geometry of observation was reconstructed from Hayabusa2 SPICE kernels available on the JAXA DARTS and the PDS NAIF (Navigation and Ancillary Information Facility) Node \citep{https://doi.org/10.17597/isas.darts/hyb2-00600}. The SP photometrically-corrected data cube of the Moon were derived in \cite{Yokota_2011} and is available on the website of the JAXA Lunar and Planetary Exploration Data Analysis (JLPEDA) group: \url{https://archive.jlpeda.isas.jaxa.jp/pub/product/moon-selene-sp/}. The disk-integrated lunar irradiance from Terra orbit for ROLO correction was computed thanks to the code developed by \cite{Kouyama_2016} and available on the GitHub repository: \url{https://github.com/TKouyama/SP_LunarCal_example/tree/main/util/SP_LunarCal_Expand_example}. The M3 L2 data were available as PDS3 volume CH1M3\_0004 in PDS3 dataset CH1-ORB-L-M3-4-L2-REFLECTANCE-V1.0 \citep{https://doi.org/10.17189/1520414} which were directly downloaded from NASA PDS Imaging Node: \url{https://planetarydata.jpl.nasa.gov/img/data/m3/CH1M3_0004/}. The photometrically-corrected M$^3$ spectral maps (raw and interpolated) prepared for this study from L2 data are freely available on the following Zenodo repository: \url{TBD}. Photometrically-corrected NIRS3 spectra (L3A) and calibrations files will be archived on the NASA PDS and JAXA DARTS.

\begin{acknowledgments}
We are grateful to the entire Hayabusa2 team for making the rendezvous with Ryugu possible. This work was supported by the JAXA Hayabusa2$\sharp$ International Visibility Enhancement Project. We would like to thank the two anonymous referees for their constructive feedback which greatly improve this manuscript.
\end{acknowledgments}





\appendix

\section{Wavelength dependency of the correction factors for data obtained between TD1 and TD2, and after TD2} \label{appendix:scale_factor_wavelength}
\begin{figure}
\centering
  \resizebox{\hsize}{!}{\includegraphics{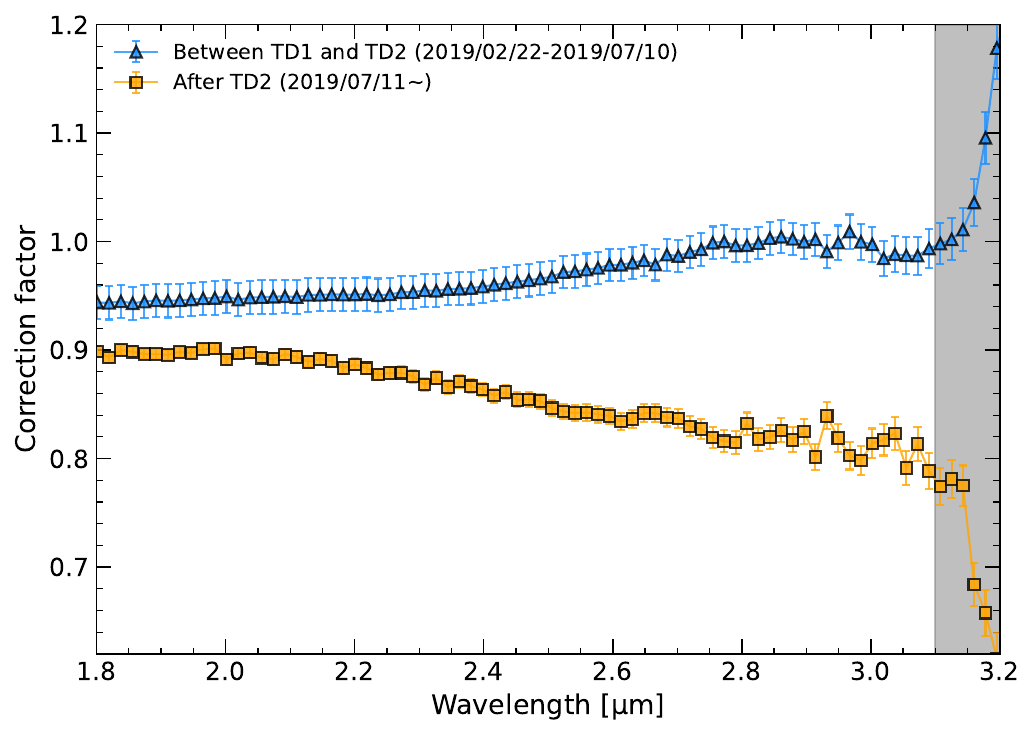}}
  \caption{Relative correction factor as a function of the wavelength between 1.8 $\mu$m and 3.2 $\mu$m. The correction factor is computed in order that data obtained after TD1 can be matched with the reflectance level of the data before TD1, which contains original and not altered calibration. The gray area corresponds to the wavelength region where the spectra exhibit calibration and/or thermal tail removal residuals.}
  \label{fig:scale_factors_vs_wavelength}
\end{figure}

In Sect. \ref{sect:relative_calib}, we described the method and the results of the investigation of the systematic radiance factor values differences between TD1 and TD2, and after TD2, compared to the data obtained before TD1. In Fig. \ref{fig:scale_factors_vs_wavelength}, we present in more details the discussed variation of the correction factors within the wavelength range of NIRS3. \par
\begin{figure*}
\centering
  \resizebox{\hsize}{!}{\includegraphics{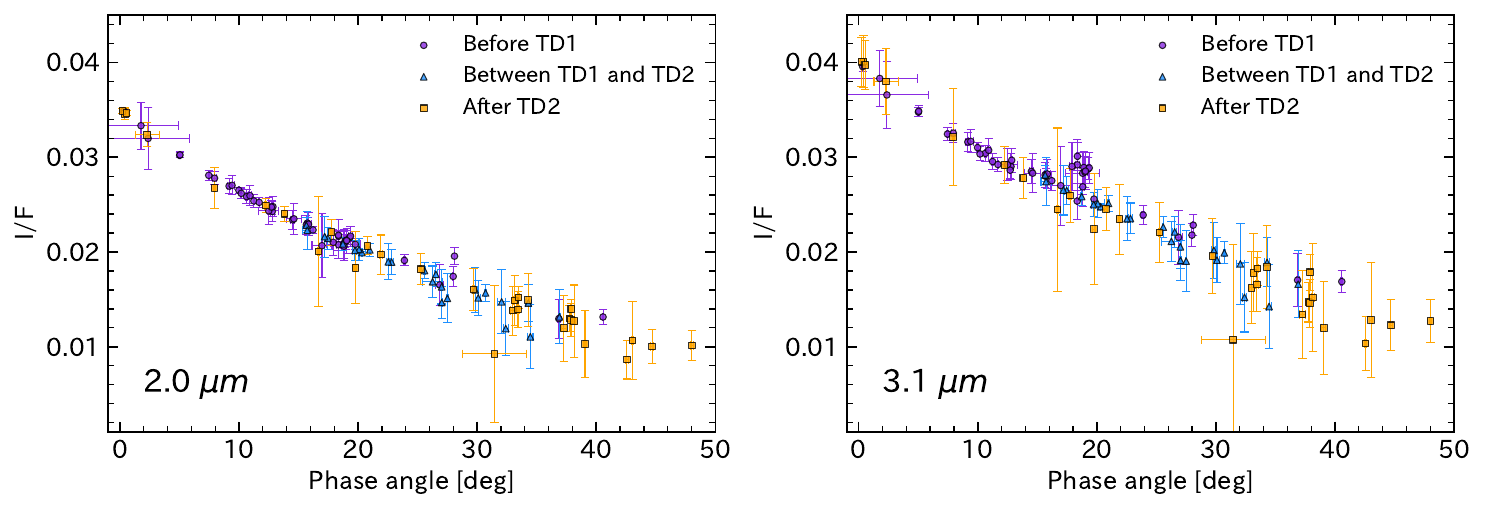}}
  \caption{Ryugu phase curves from NIRS3 data before TD1 (purple circles), between TD1 and TD2 (blue triangles), and after TD2 (orange squares) after applying the results of our updated calibration; (left) at 2 $\mu m$ (right) at 3.1 $\mu m$.}
  \label{fig:phase_curve_2selected_wavelengths}
\end{figure*}

The derived uncertainties on the correction factor derived from uncertainties of the phase curve is roughly constant with wavelength for data obtained between TD1 and TD2. On the other hand, the data acquired after TD2 exhibit increasing uncertainties with increasing wavelength. This effect is clearly visible on the data themselves and then on the derived phase curve for each mission phases (Fig. \ref{fig:phase_curve_2selected_wavelengths}). Uncertainties at the longer wavelengths are particularly significant for data after TD2.

\section{Photometric models} \label{appendix: models}
In this Appendix, we briefly summarized the photometric models used in this study.
\subsection{Hapke model} \label{appendix:hapke}
We used the Hapke IMSA model \citep{Hapke_2012} with the porosity correction, shadowing function, and shadow-hiding opposition effects. The choice of the model was based on previous photometric studies of Ryugu from \cite{Tatsumi_2020} using ONC-T data and from \cite{Domingue_2021} using NIRS3. In addition, this version of the model was used to produced level 2e ONC-T data ("Derived Photometrically Corrected Reflectance Image Data"). For this was, we used a slightly different version compared to \cite{Tatsumi_2020} and \cite{Domingue_2021}, because we took into account the porosity correction factor $K$. The radiance factor is given by the following equation:
\begin{multline}
    \frac{I}{F} = \frac{K \omega}{4} \frac{\mu_{0,e}}{\mu_{0,e}+\mu_{e}} S(i,e,\alpha,\bar{\theta}) \\ \times \left\{P_{hg}(\alpha,g) \left[1+B_{sh}(\alpha, B_{sh,0}, h_{sh})\right] + M\left(\frac{\mu_{0,e}}{K}, \frac{\mu_e}{K}, \omega \right)\right\}
\end{multline}
where $\mu_0$ and $\mu$ are respectively the cosine of the effective incidence and emergence angles, and $\omega$ is the single-scattering albedo. \newline
The $B_{sh}$ function describes the shadow-hiding opposition effect (SHOE):
\begin{equation} \label{eq:SHOE}
    B_{sh}(\alpha,B_{sh,0},h_{sh}) = \frac{B_{sh,0}}{1 + \frac{\tan \alpha/2}{h_{sh}}}
\end{equation}
where $B_{sh,0}$ is the amplitude of the SHOE, and $h_{sh}$ is the half-width of the SHOE.\newline
M is the multiple scattering function given by the following equation:
\begin{equation}
    M\left(\mu_0, \mu, \omega \right) = H\left(\mu_0,\omega \right) H\left(\mu, \omega \right) - 1   
\end{equation}
where $H$ is the Hapke's second-order approximation of the Chandrasekhar's function \citep{Hapke_2012}. $S$ is the shadowing function and is described in detail in \cite{Hapke_2012}.\newline
K is the porosity factor. We used the approximation from \cite{Helfenstein_2011} which makes the porosity factor dependent on the half-width of the SHOE:
\begin{equation}
    K = 1.069 + 2.109 h_{sh} + 0.577 h_{sh}^2 + 0.062 h_{sh}^3
\end{equation}
$P_{hg}$ is the Henyey-Greenstein (HG) phase function. We used both the one-term HG (1T-HG):
\begin{equation}
    P_{1T-HG}(\alpha,g) = \frac{1-g^2}{(1 + 2g\cos\alpha +g^2)^{3/2}}
\end{equation}

\subsection{LinMagLS} \label{appendix:LinMagLS}
We considered a model with a linear magnitude phase function and a Lommel-Seeliger disk-function, described by the following equation:
\begin{equation}
    \frac{I}{F} = A_{N}  \frac{2 \cos{i}}{\cos{i}+\cos{e}} 10^{-0.4 \beta \alpha},
\end{equation}
where $A_N$ is the normal albedo and $\beta$ is the phase slope parameter. These two variables are the free parameters of the model.

\subsection{LinMagAkimov} \label{appendix:LinMagAkimov}
We used again the linear magnitude phase function but now associated with an Akimov ($\eta = 1$) disk function \citep{Akimov_1988, Shkuratov_1994}:
\begin{equation}
    \frac{I}{F} = A_{N} \cdot 10^{-0.4 \beta \alpha} \cdot \cos\left(\frac{\alpha}{2}\right) \cos{\left(\frac{\pi}{\pi - \alpha}\left( l - \frac{\alpha}{2}\right)\right)} \left[ \frac{\left(\cos b\right)^{\frac{\alpha}{(\pi - \alpha)}}}{\cos l} \right],
\end{equation}
where $b$ and $l$ are the photometric latitude and the photometric longitude, respectively. These two variables are directly given by the illumination angles \citep{Shkuratov_2011}. The free parameters are again the normal albedo and the phase slope parameter.

\subsection{ExpMcEwen} \label{appendix:ExpMcEwen}
We also considered an exponential phase function combined with a McEwen disk function with an exponential polynomial partition function:
\begin{equation}
    \frac{I}{F} = A_{N} \cdot \left[L(\alpha) \frac{2 \cos{i}}{\cos{i}+\cos{e}} + \left(1 - L\left(\alpha\right)\right) \cos i\right] \cdot e^{\beta \alpha + \gamma \alpha^{2} + \delta \alpha^{3}},
\end{equation}
where $L(\alpha)$ is the exponential polynomial partition function described by:
\begin{equation}
    L(\alpha) = e^{\epsilon \alpha + \zeta \alpha^{2} + \eta \alpha^{3}}.
\end{equation}
This model has 7 free parameters: the normal albedo $A_N$ and the 6 polynomial coefficients $\beta, \gamma, \delta, \epsilon, \zeta, \eta$.

\subsection{ROLO-LS} \label{appendix:ROLO-LS}
The last model we used in this work is a ROLO phase function associated with a Lommel-Seeliger disk function:
\begin{equation}
    \frac{I}{F} = \frac{2 \cos{i}}{\cos{i}+\cos{e}} \left[C_{0} e^{-C_{1}} + A_{0} + A_{1} \alpha + A_{2} \alpha^{2} + A_{3} \alpha^{3} + A_{4} \alpha^{4} \right],
\end{equation}
where $C_{0}$, $C_{1}$, $A_{0}$, $A_{1}$, $A_{2}$, $A_{3}$, $A_{4}$ are the free parameters of the model.

\subsection{SP model}
Photometric studies of the lunar SP data were performed by the SP model defined by \cite{Yokota_2011}. This model uses a McEwen disk function with a third order polynomial limb darkening function:
\begin{equation}
    L_{SP}(\alpha) = 1 + c_{1}\alpha + c_{2}\alpha^{2} + c_{3}\alpha^{3}
\end{equation}
where $c_{1}, c_{2}$ and $c_{3}$ are coefficients determined by \cite{McEwen_1996} and used in \cite{Kouyama_2016}.
The phase function is defined as:
\begin{equation}
    f_{SP}(\alpha) = P_{2T-HG}(\alpha, g, c) \left[1 + B_{sh}(\alpha, B_{sh,0}, h_{sh})\right]
\end{equation}
where the SHOE function $B_{sh}$ is defined in Eq. \ref{eq:SHOE} and:
\begin{multline}
        P_{2T-HG}(\alpha, g, c) = \frac{1+c}{2}\frac{1-g^2}{(1 - 2g\cos\alpha +g^2)^{3/2}} \\ + \frac{1-c}{2}\frac{1-g^2}{(1 + 2g\cos\alpha +g^2)^{3/2}}
\end{multline}
Note that the SP model used a slightly different version of the 1T-HG function compared to the one we used in the Hapke model, resulting in an opposite sign of the asymmetry parameter $g$. \par
In summary, the SP model radiance factor is given by the following formula:
\begin{equation}
    \left.\frac{I}{F}\right|_{SP} = f_{SP}(\alpha) \left[L(\alpha) \frac{2 \cos{i}}{\cos{i}+\cos{e}} + \left(1 - L\left(\alpha\right)\right) \cos i\right] 
\end{equation}

\subsection{M$^{3}$ model}
M$^{3}$ model is a simple Lommel-Seeliger disk function associated with a degree 4 polynomial phase function. The model is described in details in \cite{Besse_2013}. The reflectance can be computed as follows:
\begin{equation}
    \left.\frac{I}{F}\right|_{M^3} = \frac{\cos{i}}{\cos{i} + \cos{e}} \left[ A_{0} + A_{1} \alpha + A_{2} \alpha^{2} + A_{3} \alpha^{3} + A_{4} \alpha^{4} \right]
\end{equation}

\section{Interpolation of the M$^3$ data} \label{appendix:interpolation_M3}
\begin{figure}
\centering
  \resizebox{\hsize}{!}{\includegraphics{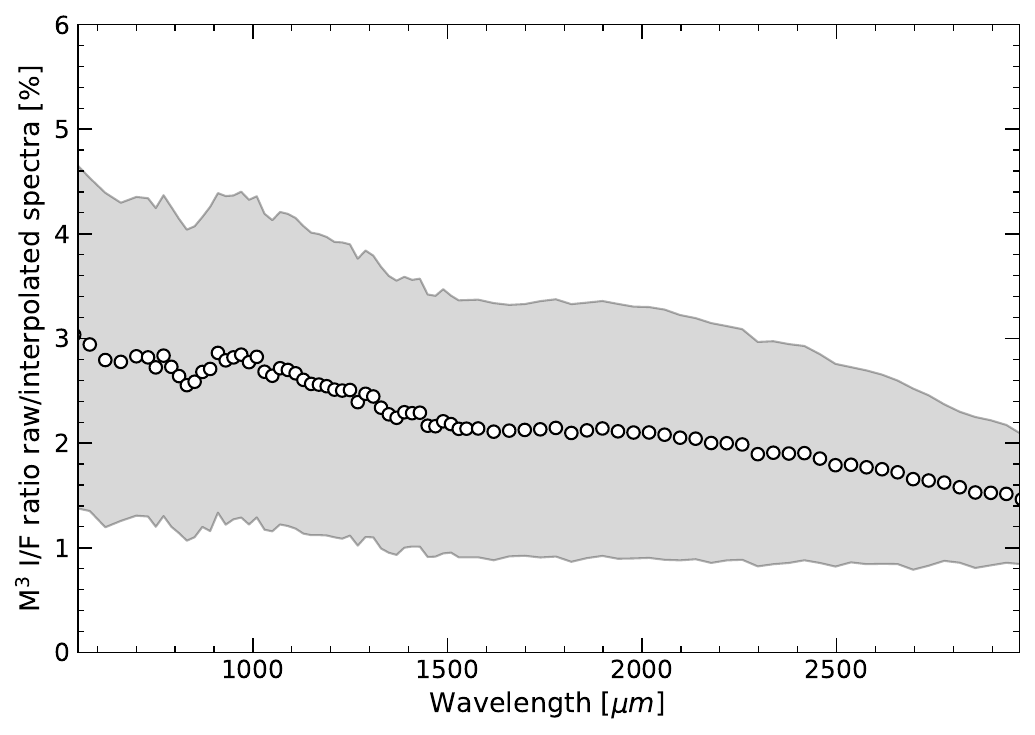}}
  \caption{Average ratio of raw and interpolated radiance factor from M$^3$ data [in \%]. The average is computed for the 45 NIRS3 lunar spectra considered. The gray shaded area corresponds to the standard deviation.}
  \label{fig:M3_interpolation_err_estimation}
\end{figure}

M$^{3}$ observations do not cover the full surface of the Moon (Fig. \ref{fig:moon_nirs3_footprint}). To integrate the lunar radiance factor from M$^3$ for each NIRS3 footprint, we explored two different approaches. The first one consists in neglecting all pixels with no radiance factor values, hence consider only valid radiance factor pixels. The second method involves a 2D interpolation for each wavelength channel and integrating afterward the radiance factor within each footprint. To evaluate the spectroscopic differences resulting from the two methods, we computed the ratio of the M$^{3}$ spectrum for each NIRS3 footprint. The result of this investigation is given in Fig. \ref{fig:M3_interpolation_err_estimation}. The ratio depends largely on the region covered by the footprint: the footprint with more unobserved surface (by M$^{3}$) introduce more uncertainties. The ratio shows also a dependency with the wavelength: at 550 nm the ratio is 3\%, while at 2950 nm it is divided by two with a value of 1.5\%. 

\section{Details for the derivation of the correction factor from SP and M$^3$ data} \label{appendix:correction_factor_SP_M3}
\begin{figure}
\centering
  \resizebox{\hsize}{!}{\includegraphics{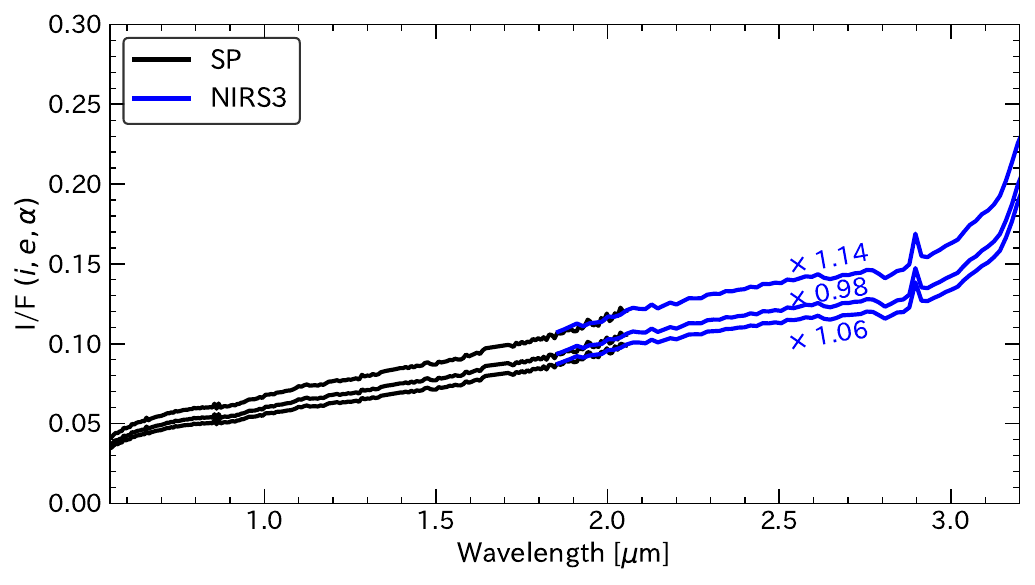}}
  \caption{Example of three comparisons of spectra from SP data (black solid line) and NIRS3. SP is reprojected to the NIRS3 geometry before the comparison. NIRS3 spectra (blue solid line) are corrected by their individual best fit scale factor.}
  \label{fig:SP_correction_example}
\end{figure}

\begin{figure}
\centering
  \resizebox{\hsize}{!}{\includegraphics{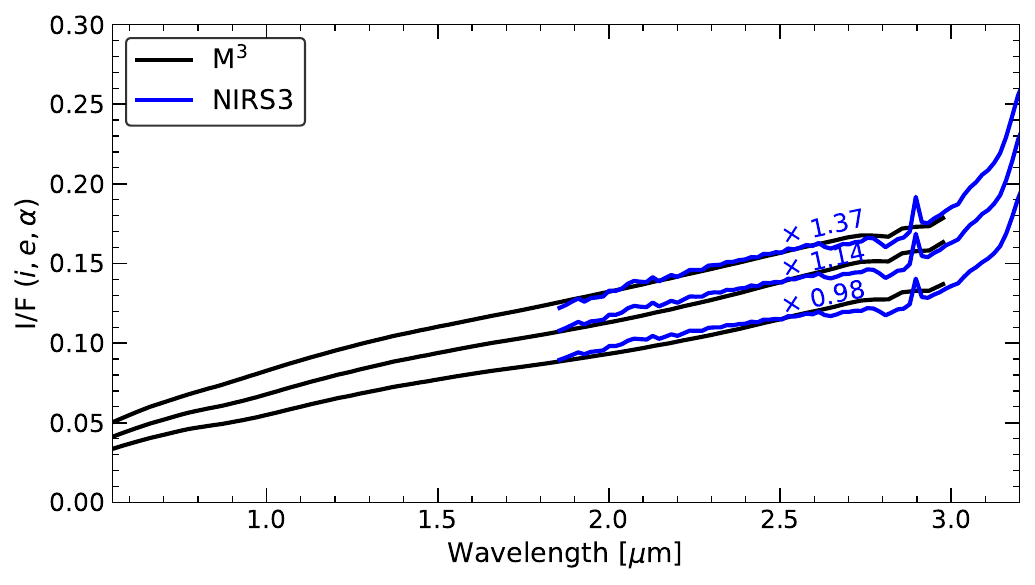}}
  \caption{Example of three comparisons of spectra from M$^3$ data (black solid line) and NIRS3. M$^{3}$ is reprojected to the NIRS3 geometry before the comparison. NIRS3 spectra (blue solid line) are corrected by their individual best fit scale factor.}
  \label{fig:M3_correction_example}
\end{figure}

To derive the final scale factor from SP and M$^3$ data, we considered a similar method between the two instruments but with slightly different parameters because of their different wavelength range. In both cases, we fitted the overlapping wavelength region between SP or M$^{3}$, and NIRS3 (1820 -- 2048 nm for SP and 1850 -- 1976 nm for M$^3$) with a least-square algorithm. The wavelength domains were chosen to obtain the largest wavelength range but avoiding as possible potential calibration issue residuals. This is why we avoided using the first channels of the effective wavelength range of NIRS3. After extracting the overlap region, to perform the fitting procedure, we first interpolate the projected spectrum from SP or M$^3$ model to the wavelength grid of the NIRS3. We obtained an individual scale factor for each of the 45 lunar NIRS3 spectra. The final scale factor is computed from the average of the 45 scale factors and the associated errors by the standard deviation value. Examples of corrected NIRS3 spectra using SP and M$^3$ are shown in Figs. \ref{fig:SP_correction_example} and \ref{fig:M3_correction_example}, respectively. \par
\begin{figure}
\centering
  \resizebox{\hsize}{!}{\includegraphics{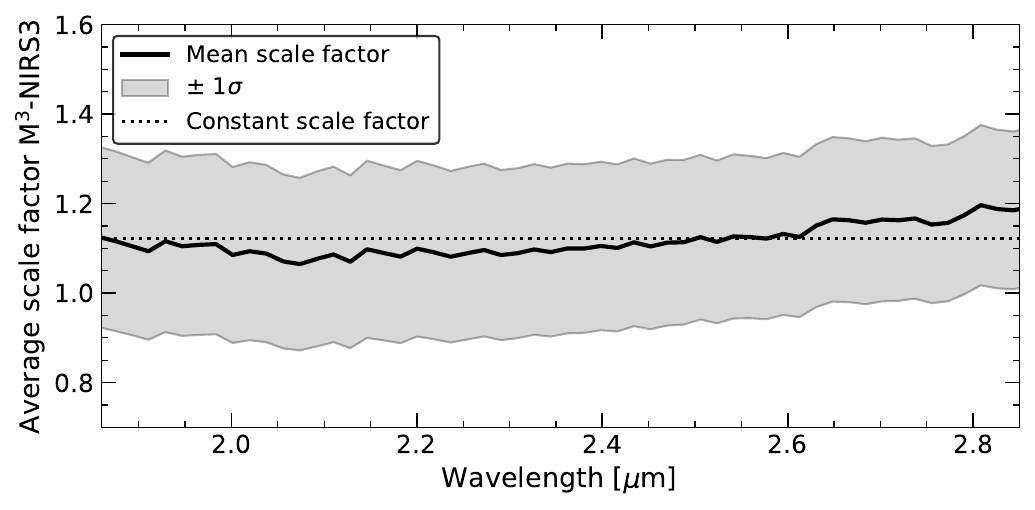}}
  \caption{Average of the scale factors used to match M$^3$ model spectra with the 45 NIRS3 spectra for each wavelength within 1.85 -- 2.85 $\mu m$ range (black solid line). The gray shaded area represents the standard deviation of the scale factors. The black dotted line is the result when considering a constant scale factor.}
  \label{fig:M3_NIRS3_scale_factore_wvl}
\end{figure}

The derivation of the wavelength-dependent M$^3$-NIRS3 scale factor follows the exact same method, but with a fit for each NIRS3 wavelength channel instead of the full overlapping wavelength range. The result of this investigation is given in Fig. \ref{fig:M3_NIRS3_scale_factore_wvl}. The value of the constant scale factor is within $\pm5$\% compared to the mean wavelength-dependent scale factor.  

\section{Additional data for the comparison between ONC-T and NIRS3 spectra} \label{appendix:NIRS3-ONC_comp}
We presented in this Appendix more details about the comparison between the spectra obtained in the visible wavelength range with the ONC-T multi-band camera and the near-infrared spectra from NIRS3. \par
\begin{figure}
\centering
  \resizebox{\hsize}{!}{\includegraphics{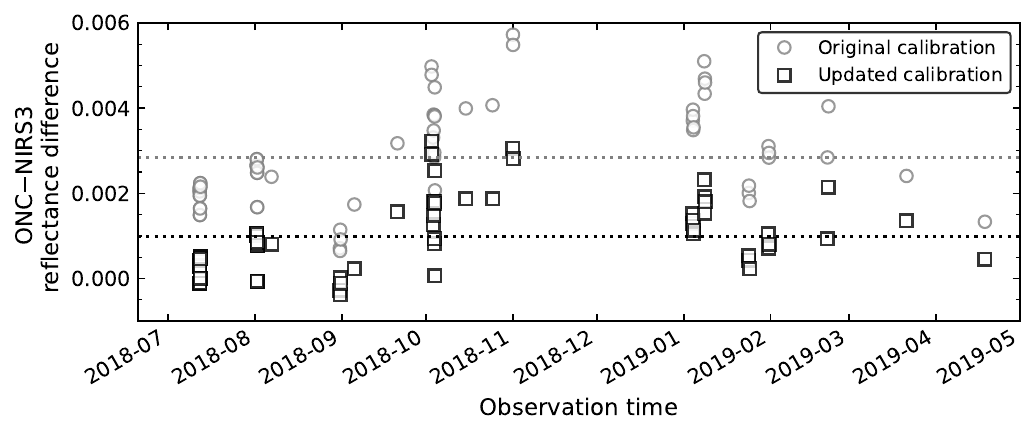}}
  \caption{Comparison between ONC-T and NIRS3 reflectance level for image/spectrum acquired at a similar time ($\pm$ 2 seconds), and for the same region of the Ryugu surface. The reflectance level is computed from the average of the reflectance at \textit{x} (0.86 $\mu m$) and \textit{p} (0.95 $\mu m$) filters for ONC-T, and from the average of the five first effective wavelength channels of NIRS3. The dotted lines represent the mean reflectance differences for the NIRS3 original calibration (in gray) and for the NIRS3 updated calibration proposed from this work (in black).}
  \label{fig:ONC_NIRS3_comp_all}
\end{figure}

We compared the reflectance of Ryugu for the same region of the surface acquired with the same observation acquisition, and especially with a quasi-identical acquisition time. For all these images/spectra, we computed the average reflectance level of ONC-T of the three $w, x, p$ filters (band center at 700 nm, 857 nm, and 945 nm, respectively); and the reflectance level for NIRS3 of the five channels at the shortest wavelength. We used only the channels at wavelength longer than 1850 nm because, even if the effective wavelength range starts at 1800 nm, the region between 1800 -- 1850 nm exhibits calibration residuals. Fig. \ref{fig:ONC_NIRS3_comp_all} presents the difference between these ONC-T and NIRS3 average reflectance levels as a function of acquisition date. Before applying the updated calibration, the NIRS3 spectra were systematically darker than ONC-T spectra, showing an average reflectance of about 0.003. This is not expected because ground-based observations from \cite{Moskovitz_2013} and \cite{LeCorre_2018} show that Ryugu spectrum is slightly red in the 945 -- 1800 nm range not covered by ONC-T and NIRS3 observations. The updated calibration improves this issue, with an average reflectance difference of approximately 0.001 between the two instruments.

\begin{figure}
\centering
  \resizebox{\hsize}{!}{\includegraphics{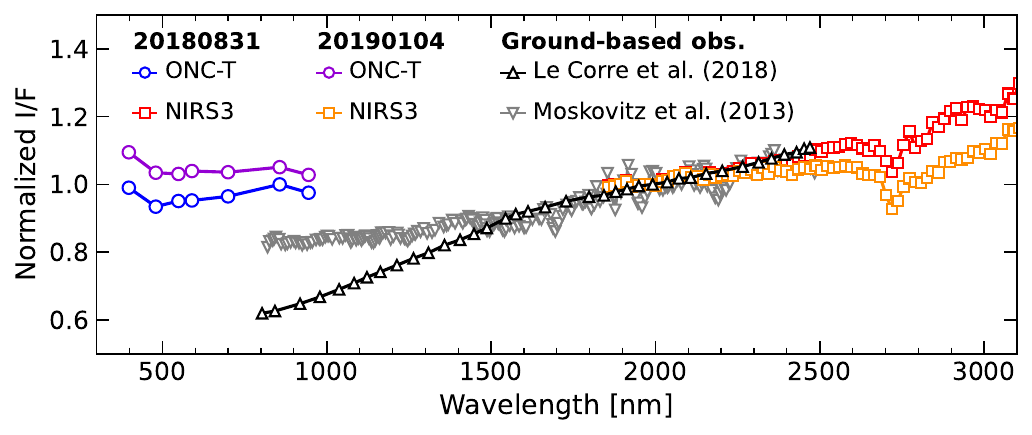}}
  \caption{Comparison between ONC-T, NIRS3 reflectance spectra (after applying the updated calibration) acquired at a similar time and for the same region of the Ryugu surface, and ground-based observations from \cite{Moskovitz_2013} and \cite{LeCorre_2018}. \cite{Moskovitz_2013} spectrum was smoothed using a Savitzky-Golay filter with a polynomial of degree 3 and a window of 5 points. All spectra are normalized at 2 $\mu m$.}
  \label{fig:ONC_NIRS3_ground-based_comp}
\end{figure}

Fig. \ref{fig:ONC_NIRS3_ground-based_comp} provides an example of comparison between ground-based observations and Hayabusa2 observations of Ryugu. The continuous ONC-T/NIRS3 spectra show a good agreement compared to ground-based observations. Observations of an irregular bodies such as Ryugu is highly dependent on the observation geometry, which may explain both the differences observed between the two ground-based observations and between ground-based and remote-sensing.

\section{Observation geometries of the proximity phase dataset (2018/06 -- 2019/11)} \label{appendix:obs_conditions}
\begin{figure}
\centering
  \resizebox{\hsize}{!}{\includegraphics{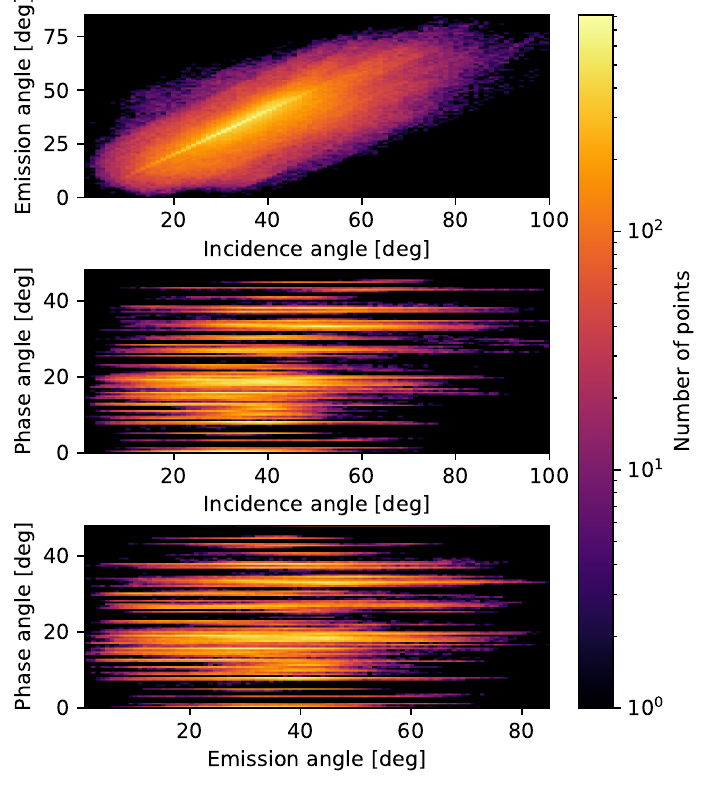}}
  \caption{Distribution of the illumination angles (incidence, emission, and phase) for the entire NIRS3 proximity phase dataset.}
  \label{fig:ill_conditions_bin}
\end{figure}

The photometric properties were derived from the entire set of NIRS3 data obtained during the Hayabusa2 proximity phase from June 2018 to November 2019. During this period, the spacecraft remains in the home position -- which corresponds to a distance of approximately 20 km from the Ryugu surface. However, it changes its altitude for specific operations, such as TD rehearsal, TD1 and TD2. These orbit changes result in significant variations in the observation conditions during this phase, particularly the spatial resolution and the illumination and viewing conditions. (Fig. \ref{fig:ill_conditions_bin}). \par
Incidence angle ranges between 1.2$\degree$ and 108.4$\degree$, emission angle between 0.6$\degree$ and 84.9$\degree$, and phase angle between 0.13$\degree$ and 48.2$\degree$. However, this distribution of illumination and viewing conditions are not exactly homogeneous for the data before and after the TD operations. This reinforces the necessity of using all data from proximity phase to have a wider ($i,e,\alpha$) distribution and finally a better estimation of the Ryugu's phase function.

\begin{figure}
\centering
  \resizebox{\hsize}{!}{\includegraphics{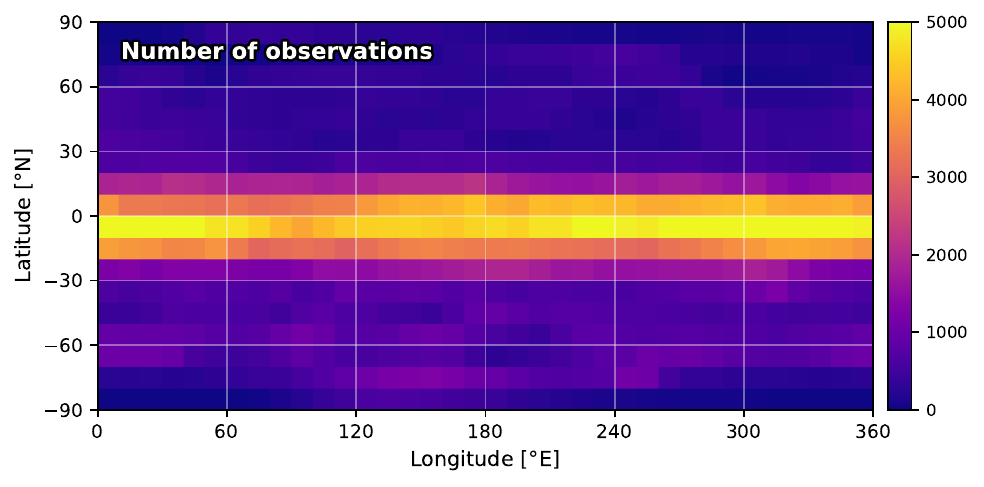}}
  \caption{Maps of the number of observations per $10\degree \times 10\degree$ grid element.}
  \label{fig:footprint_count}
\end{figure}

Most of the observations were obtained with a spatial resolution better than 40 meters, which corresponds to the typical resolution at home position \citep{Iwata_2017}. However, it is also important to note that the NIRS3 observations were not distributed homogeneously across the surface. As shown in Fig. \ref{fig:footprint_count}, most of the observations were obtained along the equatorial region ($|\text{lat}|<20\degree$). Therefore, the solar phase angle distribution is also highly heterogeneous: while the equatorial regions have phase angle covering from the opposition to more than 45$\degree$, the higher latitudes have a poor phase angle distribution (Fig. \ref{fig:phase_minmax}), making difficult the photometric studies of these regions (see Appendix \ref{appendix:maps}). In addition, the spectra obtained with the highest spatial resolution were obtained only in the equatorial regions.

\begin{figure*}
\centering
  \resizebox{\hsize}{!}{\includegraphics{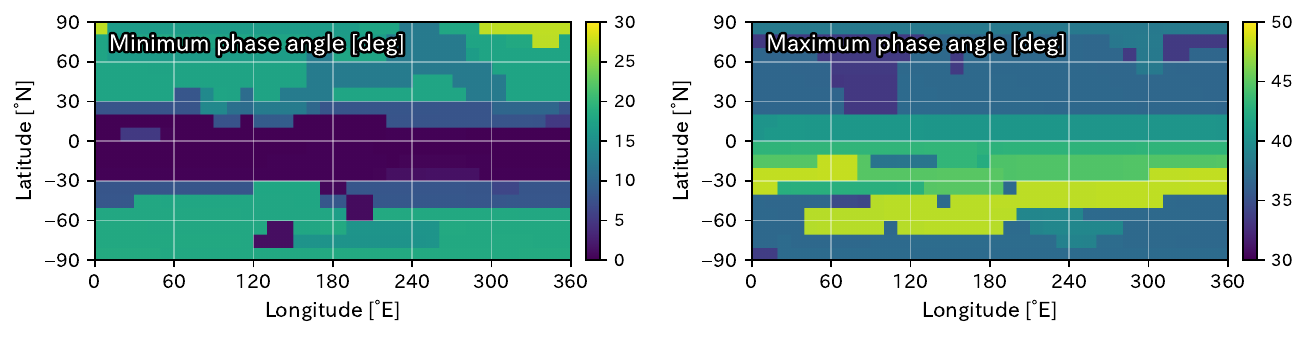}}
  \caption{Maps of the (left) minimum and (right) maximum phase angle per $10\degree \times 10\degree$ grid element.}
  \label{fig:phase_minmax}
\end{figure*}

\section{Fit comparison between the various photometric models} \label{appendix:rms}
\begin{figure}
\centering
  \resizebox{\hsize}{!}{\includegraphics{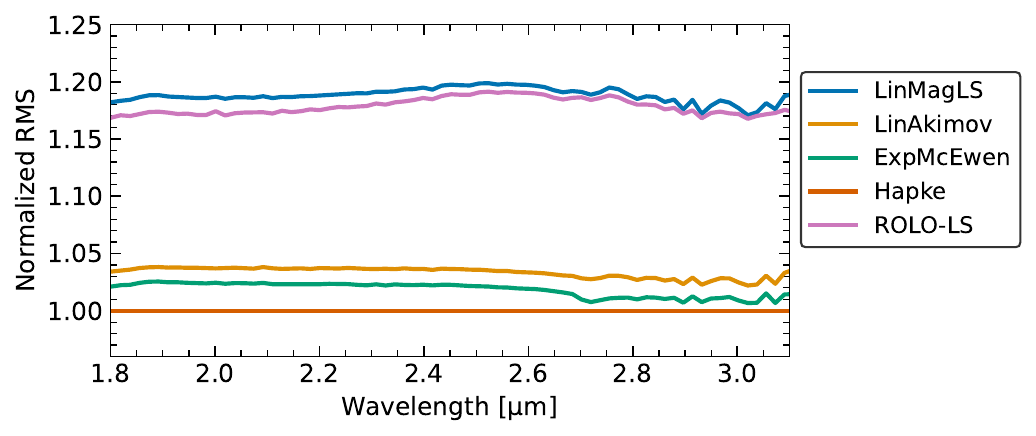}}
  \caption{RMS normalized to Hapke RMS values as a function of wavelength.}
  \label{fig:RMS_diff_with_wvl}
\end{figure}

In order to assess the quality of the fit provided by the photometric models used in this work, we computed the RMS over the NIRS3 wavelength range. Fig. \ref{fig:RMS_diff_with_wvl} shows the RMS values normalized to the one of Hapke. As discussed in Sect. \ref{sec:choice_model}, Hapke provides the best fit among the various models. The LinAkimov and ExpMcEwen provided also good fit of the data, but the number of parameters are higher than Hapke and make them less suitable. ROLO-LS and LinMagLS provides also relatively interesting fit for the phase angle larger than 10$\degree$, but the fact that they lack an opposition term did not make them usable for this dataset, with the final goal of providing good photometric correction of the NIRS3 for all illumination and viewing conditions.

\section{Sensitivity of the retrieved Hapke parameters to the macroscopic roughness} \label{appendix:sensitivity}
Because NIRS3 did not observe Ryugu at high phase angles, the macroscopic roughness parameter ($\bar{\theta}$) is poorly constrained. In the Hapke fitting procedure, we solved this parameter alone at the penultimate step, at fixed $w$, $g$, $B_{sh,0}$, and $h_{sh}$. The very small spectral dependency we observed for $\bar{\theta}$ is therefore likely artificial and due to the strong constraints we put on the other parameters before determining $\bar{\theta}$. As this could affect the retrieved $w$ and $g$ spectra, we tested the sensitivity of the fit to this assumption. \par
To perform this sensitivity analysis, we repeated the final refinement step ($w$, $g$, $B_{sh,0}$, $h_{sh}$ free) at each wavelength, each time holding $\bar{\theta}$ fixed at $\bar{\theta}_{\mathrm{best}} + \Delta\bar{\theta}$, with $\Delta\bar{\theta} \in \{-5^\circ, -2.5^\circ, 0^\circ, +2.5^\circ, +5^\circ\}$. From the five set of Hapke parameters, we modeled the resulting radiance factor spectra at the standard geometry $i=30^\circ$, $e=0^\circ$, $\alpha=30^\circ$. The depth of the 2.72 $\mu$m absorption band was then measured on each spectrum by removing the linear continuum. Across the $\pm5^\circ$ range in $\bar{\theta}$, both the $w$ and $g$ spectra vary mostly in absolute values, the slight wavelength-dependent variations are likely noise. Across all valid wavelength channels, $w$ shows a mean spread of 0.0038 and $g$ exhibits a value of 0.0218. For both parameters, these variations are about four times larger than the typical fit uncertainties (see Table \ref{tab:hapke_params_global}). Similarly, the 2.72 $\mu$m band depth do not change with the $\bar{\theta}$ value, with a maximum variation smaller than 0.10\%. \par
The absolute value of $\bar{\theta}$ do not modify the spectral properties of the photometric parameters and the retrieved 2.72 $\mu$m band depth, indicating that the fixed-$\bar{\theta}$ assumption in the third round of the fitting procedure does not introduce bias in our spectral interpretation of the hydroxyl feature.

\begin{figure}
\centering
  \resizebox{\hsize}{!}{\includegraphics{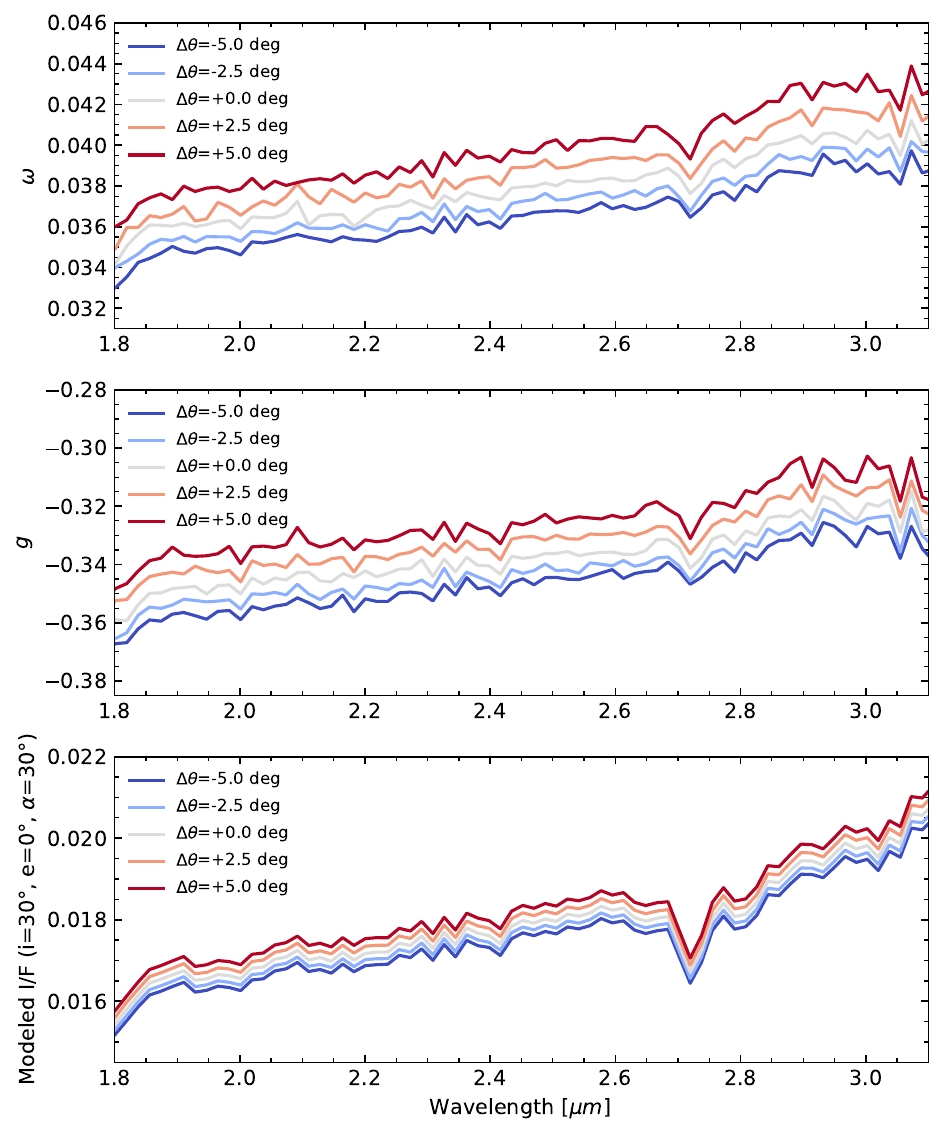}}
  \caption{RMS normalized to Hapke RMS values as a function of wavelength.}
  \label{fig:sensitivity}
\end{figure}

\section{Relationship between phase reddening and spatial resolution} \label{appendix:phase_reddening_resolution}
\begin{figure}
\centering
  \resizebox{\hsize}{!}{\includegraphics{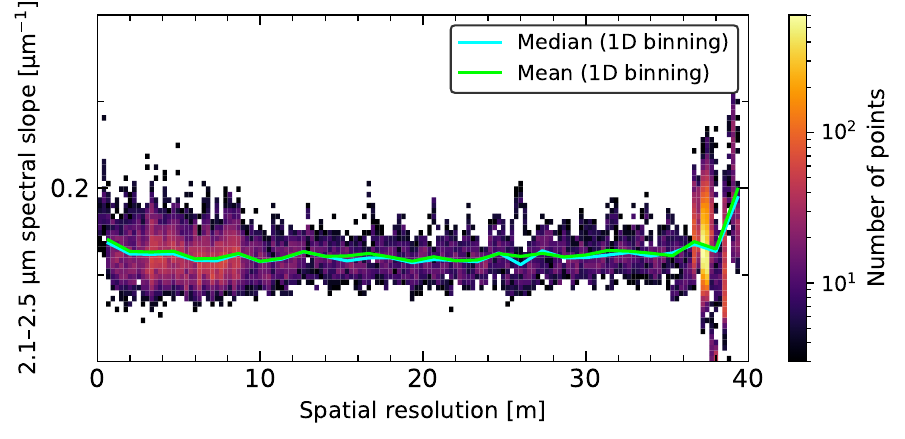}}
  \caption{Spectral slope (2.1 -- 2.5 $\mu m$) distribution as a function of the NIRS3 spatial resolution [m]. Data are restricted to similar geometric conditions ($20\degree <\alpha < 30\degree, i < 50\degree, e < 50 \degree$). The data were binned in a 2D slope-resolution grid using 150 $\times$ 150 bins. The colorbar corresponds to the number of points per bin. We considered only bins containing at least 3 points, others are considered empty. The cyan and lime solid lines represent the median and mean value from each resolution bin, respectively.}
  \label{fig:resolved_slope_resolution_bin}
\end{figure}

Because phase reddening has been demonstrated to results from variations of the physical properties such as grain size and/or roughness \citep{Schroder_2014}, phase reddening estimation may be slightly different for observations acquired with important differences in spatial resolution. The spatial resolution for a given observation is retrieved from the field of view (FOV = 0.11$\degree$, \citealt{Iwata_2017}) and the altitude over Ryugu surface of the spacecraft. We first investigated the evolution of the spectral slope with the resolution after applying the updated NIRS3 calibration (Fig. \ref{fig:resolved_slope_resolution_bin}). The spectral slope is constant ($\sim$0.16 -- 0.17 $\mu$m$^{-1}$) across the entire range of spatial resolutions considered (0 -- 40 meters), with variations of the mean typically below 4\%. The observed scattering of the spectral slope values is mainly caused by variations in emission and incidence angles.

\begin{figure}
\centering
  \resizebox{\hsize}{!}{\includegraphics{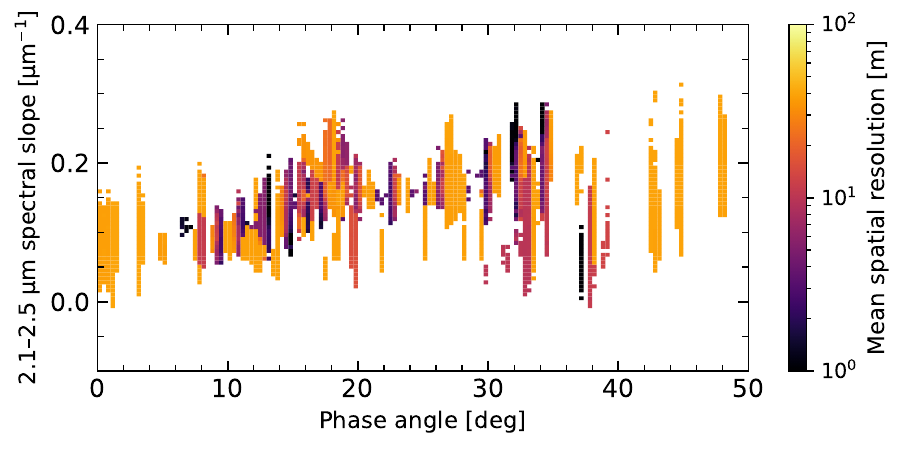}}
  \caption{Spectral slope (2.1 -- 2.5 $\mu m$) distribution as a function of the solar phase angle [deg]. The data were binned in a 2D slope-resolution grid using 150 $\times$ 150 bins. The colorbar corresponds to the mean spatial resolution per bin. We considered only bins containing at least 5 points, others are considered empty.}
  \label{fig:resolved_reddening_resolution_bin}
\end{figure}

Similarly, the phase reddening (i.e., the evolution of spectral slope with phase angle) do not show variations with the spatial resolution. NIRS3 observations acquired at different spatial scale (within two orders of magnitude variations) have a similar spectral slope (Fig. \ref{fig:resolved_reddening_resolution_bin}).

Based on these results, all observations acquired at spatial resolutions from 0 to 40 meters are included in this study.


\section{Hapke parameter maps} \label{appendix:maps}
\begin{figure*}
\centering
  \resizebox{\hsize}{!}{\includegraphics{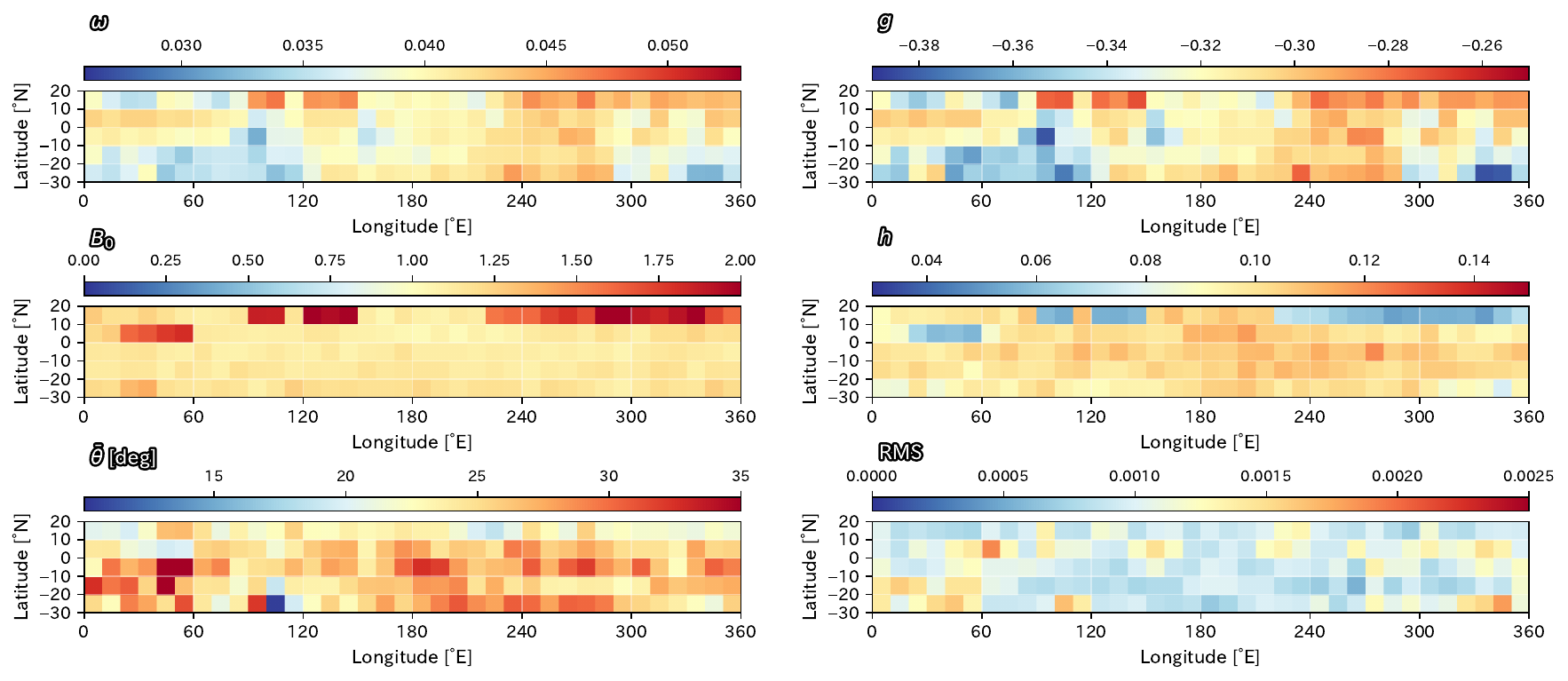}}
  \caption{Hapke parameters maps at 2.0 $\mu m$ using a $10\degree \times 10\degree$ grid at the Ryugu surface.}
  \label{fig:hapke_maps}
\end{figure*}

This Appendix presents the photometric parameter maps from Hapke model (Fig. \ref{fig:hapke_maps}). Because opposition data were only obtained for the equatorial region, we limited our regional photometric analysis to the latitudes from -20$\degree$N to 10$\degree$N. \par
The parameters $\omega$ and $g$ appears to be highly spatially correlated. An east-west dichotomy is observed for almost all parameters (Fig. \ref{fig:hapke_maps} and Table \ref{tab:east_west_dichotomy}). The single scattering albedo appears brighter in the western bulge with a 3.8\% higher value. This is consistent with albedo variations in the visible noticed by \cite{Sugita_2019} and \cite{Tatsumi_2020}. The relative modification of the SSA between eastern and western hemisphere are smaller than the one derived by \cite{Tatsumi_2020}, but consistent with the latter study which found a wavelength-dependent variation. The asymmetry parameter is higher in the western hemisphere of about 3.5\%, meaning that the eastern hemisphere has a slightly stronger backscattering. For $B_{sh,0}$, the variations between the east and west hemispheres is 5.4\%. The eastern hemisphere exhibits a stronger opposition effect. The half-width of the opposition effect is larger (+8.1\%) within the western bulge. The photometric roughness $\bar{\theta}$ is the only parameter that do not exhibit specific asymmetry between the eastern and western hemisphere. \par
A higher albedo, a less backscattered surface, a stronger opposition effect amplitude: all these parameters points out towards a less important surface roughness of the western bulge compared to the eastern hemisphere. The Hapke parameters distribution across the Ryugu surface is consistent with the phase ratio and phase reddening results (Fig. \ref{fig:ratio_reddening_maps}). 

\begin{table}[]
    \centering
    \caption{East-west dichotomy of the Hapke parameters within $-20\degree \text{N}<\text{lat}<10\degree$ \text{N}.}
    \resizebox{\hsize}{!}{
    \begingroup
    \renewcommand{\arraystretch}{1.25}
    \begin{tabular}{ccc}
    \hline
    \hline
       Hapke parameters & Eastern hemisphere & Western hemisphere \\
       \hline
       $\omega$ & 0.0391 $\pm$ 0.0003 & 0.0406 $\pm$ 0.0003 \\
       $g$ & -0.324 $\pm$ 0.002 & -0.313 $\pm$ 0.002 \\
       $B_{sh,0}$ & 1.17 $\pm$ 0.02 & 1.11 $\pm$ 0.01 \\
       $h_{sh}$ & 0.098 $\pm$ 0.001 & 0.106 $\pm$ 0.001 \\
       $\bar{\theta}$ [deg] & 26.0 $\pm$ 0.4 & 26.9 $\pm$ 0.4 \\
    \hline
    \end{tabular}
    \endgroup
    }
    \label{tab:east_west_dichotomy}
\end{table}

\bibliography{sample701}{}
\bibliographystyle{aasjournalv7}



\end{document}